\documentclass[prd,preprint,tightenlines,floatfix,showpacs,preprintnumbers,nofootinbib,eqsecnum,superscriptaddress]{revtex4}

\usepackage[dvips,final]{graphicx}
 \usepackage{amssymb}
  \usepackage{amsmath}
   \usepackage{amsfonts}
    \usepackage{epsfig}
     \usepackage{bm}
      \usepackage{multirow}
       \usepackage{tabularx}

     \newcommand{\bq}{\mbox{\boldmath $q$}}
     \newcommand{\bk}{\mbox{\boldmath $k$}}

\begin{document}

\begin{flushright}
LU TP 13-32\\
September 2013
\end{flushright}
\vspace{1cm}

\title{Search for technipions in exclusive production \\
of diphotons with large invariant masses at the LHC}

\author{Piotr Lebiedowicz}
\email{Piotr.Lebiedowicz@ifj.edu.pl} \affiliation{Institute of
Nuclear Physics PAN, PL-31-342 Cracow, Poland}

\author{Roman Pasechnik}
\email{Roman.Pasechnik@thep.lu.se} \affiliation{Department of
Astronomy and Theoretical Physics, Lund University, SE-223 62 Lund,
Sweden}

\author{Antoni Szczurek}
\email{Antoni.Szczurek@ifj.edu.pl} \affiliation{University of
Rzesz\'ow, PL-35-959 Rzesz\'ow, Poland} \affiliation{Institute of
Nuclear Physics PAN, PL-31-342 Cracow, Poland\vspace{1.0cm}}

\begin{abstract}
\vspace{0.5cm} We focus on exclusive production of neutral
technipion $\tilde \pi^0$ in $pp$ collisions at the LHC, i.e. on $p
p \to p p \tilde \pi^0$ reaction. The dependence of the cross
section on parameters of recently proposed vector-like Technicolor
model is studied. Characteristic features of the differential
distributions are discussed. For not too large technipion masses the
diphoton decay channel has the dominant branching fraction. This is
also the main reason for an enhanced production of neutral
technipions in $\gamma\gamma$-fusion reaction. We discuss potential
backgrounds of the QCD and QED origin to the $p p \to p p
(\tilde{\pi}^0 \to \gamma \gamma)$ process at large invariant
$\gamma\gamma$ masses. We conclude that compared to inclusive case
the signal-to-background ratio in the considered exclusive reaction
is vary favorable which thereby could serve as a good probe for
Technicolor dynamics searches at the LHC.
\end{abstract}

\pacs{14.80.Ec, 14.80.Bn, 12.60.Nz, 14.80.Tt, 12.60.Fr}

\maketitle

\section{Introduction}

A typical central exclusive production (CEP) process with the
signature $pp\to p + X + p$, where $X$ is a diffractive system
separated from the two very forward protons by large rapidity gaps,
is considered to be very sensitive to New Physics contributions. In
particular, it has been proposed as an alternative way of searching
for neutral Higgs bosons and SUSY particles (see
Ref.~\cite{Albrow:2010yb,FP420} for a review on the topic) due to a
reduced QCD $b\bar b$ background. In this paper, we would like to
focus on an extra interesting opportunity of making use of large
rapidity gap processes at the LHC for probing new strongly-coupled
dynamics.

For the QCD-initiated CEP processes there is a serious problem of
rather large theoretical uncertainties of the QCD diffraction
mechanism in the framework of the Durham Model (see e.g.
Ref.~\cite{Durham}). These uncertainties come from both the hard
subprocess treatment and soft $k_{\perp}$-dependent parton densities
as well as from a model-dependent gap survival probability factor
(see e.g. Refs.~\cite{Dechambre:2011py,SF,chic,LKRS10}). This
situation forces the search for various possible ways to probe the
underlying CEP QCD mechanism. In order to reduce theoretical
uncertainties, new experimental data on various exclusive production
channels are certainly required and expected to come soon from
ongoing LHC measurements. In particular, a measurement of the
exclusive dijets production at the LHC could largely reduce the
theoretical uncertainty in the Higgs boson
CEP~\cite{Dechambre:2011py}. Other measurements of exclusive heavy
quarkonia \cite{chic,LKRS10}, $\gamma\gamma$ \cite{LKRS10} and $W^+W^-$
pairs \cite{LPS13}, high-$p_\perp$ light mesons
\cite{Szczurek:2006bn,HarlandLang:2011qd}, exclusive associated
charged Higgs $H^+W^-$ \cite{EP2011} CEP, etc., can also be
important in this context.

Besides the QCD-initiated CEP processes like the exclusive Higgs and
dijet production, there are extra QED-initiated contributions coming
from $\gamma\gamma\to X$ subprocesses. Normally, these contributions
are strongly suppressed by very small fine structure constant and
therefore typically neglected compared to the QCD ones especially for not very
large invariant $X$-system masses, except for leading-order
exclusive dilepton $X\equiv l^+l^-$ production. On the other hand,
the exclusive reaction via the $\gamma\gamma$ fusion have
significantly smaller theoretical uncertainties compared to the
QCD-initiated Durham mechanism making it a very appealing option for
New Physics searches for exotic resonances which are coupled to
photons or SM gauge bosons only.

Recently, the CMS Collaboration has indicated yet unexplained
resonant $2\sigma$-signature in the $\gamma\gamma$ invariant mass
spectrum around $\sim 137$ GeV \cite{CMS-bump}. Regardless of whether
this is physical or not it is worth to search for simplest possibilities
of having an extra narrow neutral resonance decaying predominantly
into the $\gamma\gamma$ pair. These exotic light physical states,
such as technipions, are naturally predicted by a high-scale
strongly-coupled dynamics commonly referred to as Technicolor
(TC) \cite{TC,Extended-TC} (for a review, see also
Ref.~\cite{Hill:2002ap,Sannino}).

In original minimal Higgs-less TC models, the EW symmetry is broken
by techniquark condensate $\langle Q\bar Q\rangle$ and there are no
composite scalars left in the spectrum since pseudo-Goldstone
technipions appearing due to the chiral symmetry breaking at a TeV
energy scale are absorbed by the SM gauge bosons. Recently, however,
the SM Higgs boson has been discovered \cite{ATLAS,CMS} leaving
practically no room for minimal TC scenarios, and the search for
consistent alternatives incorporating new strongly-coupled dynamics,
dynamical EW symmetry breaking (EWSB) and the (elementary or
composite) Higgs boson is on the way.

Many existing dynamical EWSB scenarios, including those with walking
and topcolor dynamics, incorporate more than the minimal two flavors
of techniquarks. Such scenarios feature pseudo-scalar technipion
states that are remnants of the EWSB in models with more than one
weak techniquark doublet. Discovery of such technipions is often
considered as one the basic observational signatures of TC
\cite{Jia:2012kd,Frandsen:2012rj,Hapola:2012wi}. In extended TC
scenarios with colorless (or colored) techniquarks the technipion
can be produced via gluon-gluon and quark-antiquark fusion through a
strong technipion coupling to heavy $t,b$ quarks (or
techniquark-gluon coupling). As was shown in
Ref.~\cite{Chivukula:2011ue} (and in references therein) in such
scenarios the relatively light technipions $m_{\tilde \pi}<2m_t$ are
excluded by the SM Higgs searches at the LHC. Do we still have a
room for light ($m_{\tilde \pi}\sim 100-300$ GeV) technipions
consistent with EW and LHC precision constraints?

In this paper, we consider an alternative realization of the
dynamical EWSB ideas -- the so-called vector-like TC scenario
recently proposed and discussed in detail in
Refs.~\cite{VLTC,VLTC-DM}. This model is a successful alternative to
the standard (Extended, Walking) TC implementations which is
essentially the minimal TC extension of the SM with one (elementary
or composite) Higgs doublet and extra strongly-coupled weak doublet
of vector-like techniquarks (i.e. with two ``techni-up'' $U$ and
``techni-down'' $D$ flavors only).

The idea of vector-like (chiral-symmetric) ultraviolet completion
which is fully consistent with precision EW constraints at the
fundamental level has been realized in the framework of the gauged
linear $\sigma$-model initially developed for QCD hadron physics
\cite{Lee,LSigM,SU2LR}. In this phenomenological approach, the
spontaneous {\it global} chiral symmetry breaking in the techiquark
sector happens by means of technisigma vacuum expectation value (vev)
in the chiral-symmetric (vector-like) way
\begin{eqnarray}
SU(2)_{\rm L}\otimes SU(2)_{\rm R}\to SU(2)_{\rm V\equiv L+R}\,,
\label{CSB}
\end{eqnarray}
where the resulting unbroken chiral-symmetric subgroup $SU(2)_{\rm
V\equiv L+R}$ is then {\it gauged} and therefore describes gauge
interactions of the techniquark sector. The minimality of such a
scenario which incorporates the SM Higgs sector is provided by the
fact that one gauges {\it only} the vector part of the global chiral
symmetry. In Ref.~\cite{VLTC} it was argued that the vector-like
gauge group $SU(2)_{\rm V}$ can, in principle, be {\it identified}
with the weak isospin group $SU(2)_{\rm W}$ of the SM, i.e.
\begin{eqnarray}
SU(2)_{\rm V\equiv L+R}\equiv SU(2)_{\rm W}\,, \label{ident}
\end{eqnarray}
in the techniquark sector. Such a dynamical realization of the
chiral-gauge symmetry leads to specific properties of the
techniquark sector w.r.t. weak interactions, which thereby make it
to be very different from the chiral-nonsymmetric SM fermion
sectors. It therefore means that after the chiral symmetry breaking
in the techniquark sector the left and right components of the
original Dirac techniquark fields can interact with the SM weak
$SU(2)_{\rm W}$ gauge bosons with vector-like couplings, in
contrast to ordinary SM fermions, which interact under $SU(2)_{\rm
W}$ by means of their left-handed components only.

The resulting weak isospin symmetry $SU(2)_{\rm W}$ is broken
by means of the effective SM Higgs mechanism which
thereby gets {\it initiated} by the techniquark condensation
providing the dynamical nature of the EWSB \cite{VLTC}. In this
scenario, the additional Goldstone bosons arising from the Higgs
weak doublet are absorbed by $Z,W^\pm$ bosons in the standard way
while pseudo-Goldstone technipions from extra TC dynamics remain
physical in a full analogy with QCD hadron physics. As we will see
below these technipions can be rather light, in principle, as light
as the $W$ boson since they do not couple to ordinary quarks and
gluons and could potentially be accessible to a standard Higgs boson
searches e.g. in $\gamma\gamma$ and $\gamma Z$ decay channels. Since
the diphoton channel appears to be the most favorable channel for
such technipion searches at the LHC we wish to discuss in the
present paper also the diphoton backgrounds which turn out to
be suppressed compared to the ${\tilde \pi}^0 \to \gamma\gamma$ signal
in the exclusive production process.

In this paper, we therefore consider the exclusive production of
$\gamma\gamma$ pairs which is among one of the diffractive ``golden
channels'' for both Higgs boson and light technipion searches at the
LHC. The $pp\to p(\gamma\gamma)p$ process going through the
diffractive QCD mechanism with the $gg \to \gamma\gamma$ subprocess
naturally constitutes a background for the resonant technipion
production. The photon-photon contribution for the purely exclusive
production of low invariant mass of $\gamma \gamma$ was discussed
very recently in Ref.~\cite{ES13}. There only lepton and quark loops
have been considered. In the case of technipion production at the
LHC we are rather interested in relatively large invariant diphoton
masses $M_{\gamma \gamma} \gtrsim 100$ GeV relevant for the SM Higgs
boson searches as well. In the present paper, we shall calculate
both the QCD and QED contributions and compare them differentially
as a function of diphoton invariant mass suggesting potentially
measurable a signature of vector-like Technicolor.

\section{Technipion interactions from vector-like Technicolor}

We start from vector-like TC model setup relevant for our purposes
here. The local chiral vector-like subgroup $SU(2)_{\rm V\equiv
L+R}=SU(2)_{\rm W}$ appearing due to the spontaneous global chiral
symmetry breaking (\ref{CSB}) acts on confined elementary
techniquark sector \cite{VLTC}, i.e.
 \begin{eqnarray} \label{Tdoub}
 \tilde{Q} = \left(
      \begin{array}{c}
         U \\
         D
      \end{array}
             \right)\,,
 \end{eqnarray}
which is thus in the fundamental representation of the SM gauge
$SU(2)_{\rm W}\otimes U(1)_{\rm Y}$ group and
$SU(3)_c$-neutral at the same time. As usual, in addition we have
the initial scalar technisigma $S$ field which is the SM singlet,
and the triplet of initial (massless) technipion fields
$P_a,\,a=1,2,3$ which is the adjoint (vector) representation of
$SU(2)_{\rm W}$ (with zeroth $U(1)_{\rm Y}$ hypercharge). The linear
$\sigma$-model part of the Lagrangian responsible for the
Yukawa-type interactions of the techniquarks (\ref{Tdoub}) reads
 \begin{eqnarray} \label{Yuk}
 {\cal L}_Y^{\rm TC} = -g_{\rm TC} \bar{\tilde{Q}}(S+i\gamma_5\tau_a P_a) \tilde{Q}\,,
 \end{eqnarray}
where $\tau_a,\,a=1,2,3$ are the Pauli matrices, and effective
Yukawa coupling $g_{\rm TC}>1$. After the chiral and EW symmetries
breaking, the Yukawa terms (\ref{Yuk}) determine the strength of
technipion interactions with techniquarks as well as (pseudo)scalar
self-couplings \cite{VLTC}.

Non-local effects in gauge boson couplings to technipions and
constituent techniquarks, in general, can be incorporated via
momentum-dependent form factors. In the case of a large
techniconfinement scale $\Lambda_{\rm{TC}}\sim 0.1-1$ TeV, these
effects are strongly suppressed by large constituent masses of
techniquarks $M_{Q}\sim \Lambda_{\rm TC}$ and can be neglected to
the first approximation. Thus the vector-like gauge interactions of
$\tilde{Q}$ and $P_a$ fields with initial $U(1)_{\rm Y}$ and
$SU(2)_{\rm W}$ gauge fields $B_{\mu},\,W_{\mu}^a$, respectively,
can be introduced in the local approximation via usual EW gauge
couplings $g_{1,2}$ renormalized at the $\mu=2M_Q$ scale, i.e.
 \begin{eqnarray}
  {\cal L}_{\tilde \pi,\tilde Q} = \frac12 D_{\mu} P_a\, D^{\mu} P_a +
  i \bar{\tilde{Q}}\hat{D}\tilde{Q}\,, \label{LG}
 \end{eqnarray}
where
 \begin{eqnarray}
    && \hat{D}\tilde{Q} = \gamma^{\mu} \left( \partial_{\mu}
       - \frac{iY_Q}{2}\, g_1B_{\mu} - \frac{i}{2}\, g_2 W_{\mu}^a \tau_a
       \right)\tilde{Q}\,, \label{DQ} \\
    && \qquad D_{\mu} P_a = \partial_{\mu} P_a + g_2
\epsilon_{abc} W^b_{\mu} P_c\,, \nonumber
 \end{eqnarray}
besides that $\tilde{Q}$ is also confined under a QCD-like
$SU(N_{\rm TC})_{\rm TC}$ group. In this paper, we discuss a
particular case with the number of technicolors $N_{\rm TC}=3$.

After the EWSB, the physical Lagrangian of vector-like interactions
of techniquarks and gauge bosons $V=Z^0,\,W^{\pm},\,\gamma$ reads
\begin{eqnarray}
 L_{\bar{\tilde{Q}}\tilde{Q}V}&=&g_W^Q\,\bar{U}\gamma^{\mu}D\cdot W_{\mu}^+ +
g_W^Q\,\bar{D}\gamma^{\mu}U\cdot W_{\mu}^- \nonumber \\
&+& Z_{\mu} \sum_{Q=U,D} g_Z^Q\,\bar{f}\gamma^{\mu}f +
\sum_{Q=U,D}g_{\gamma}^Q\,\bar{f}\gamma^{\mu}A_{\mu}f\,,
\label{L-QV}
\end{eqnarray}
where technifermion couplings to vector bosons $g_{V_{1,2}}^Q$ are
\begin{eqnarray}
g_Z^Q=\frac{g}{c_W}\big(t_3^Q - q_Qs_W^2\big)\,,\quad
g_W^Q=\frac{g}{\sqrt{2}}\,,\quad g_{\gamma}^Q=e\,q_Q\,.
\label{Q-gauge}
\end{eqnarray}
Here, $s_W=\sin\theta_W$, $c_W=\cos\theta_W$, and $\theta_W$ is the
Weinberg angle, $e=g s_W$ is the electron charge, $t_3^Q$ is the
weak isospin ($t_3^U=1/2$, $t_3^D=-1/2$), $q_Q=Y_{\tilde Q}/2+t_3^Q$
is the technifermion charge. Choosing the technifermion hypercharge
to be the same as in the SM fermion sector $Y_{\tilde Q}=1/3$, we
get $q_U=2/3$ and $q_D=-1/3$. Also, the Yukawa-type interactions of
constituent techniquarks with technipions are governed by
 \begin{eqnarray}
 &&L_{\bar{\tilde{Q}}\tilde{Q}\tilde{\pi}} = - i\sqrt{2}g_{\rm TC}\,
 \tilde{\pi}^+\bar{U}\gamma_5 D - i\sqrt{2}g_{\rm TC}\,\tilde{\pi}^-\bar{D}\gamma_5 U -
 ig_{\rm TC}\,\tilde{\pi}^0(\bar{U}\gamma_5 U - \bar{D}\gamma_5 D)\,.
 \label{L-QhSpi}
\end{eqnarray}
Since we are interested here in neutral technipion couplings
in exclusive production processes, only the last two terms of the
Yukawa Lagrangian (\ref{L-QhSpi}) will be used. Finally, Born-level
interactions of technipions with gauge bosons are defined as follows
\begin{eqnarray}
L_{\tilde{\pi}\tilde{\pi}V} &=&
 ig_2{W^{\mu}}^+ \cdot (\tilde{\pi}^0\tilde{\pi}_{,\mu}^- -
\tilde{\pi}^-\tilde{\pi}_{,\mu}^0) +
 ig_2{W^{\mu}}^- \cdot (\tilde{\pi}^+\tilde{\pi}_{,\mu}^0 -
\tilde{\pi}^0\tilde{\pi}_{,\mu}^+) \nonumber \\
 &+& ig_2(c_W Z_{\mu} + s_W A_{\mu})\cdot
 (\tilde{\pi}^-\tilde{\pi}_{,\mu}^+ -
 \tilde{\pi}^+\tilde{\pi}_{,\mu}^-) \nonumber \\
 &+& g_2^2\, W_{\mu}^+ {W^{\mu}}^- \cdot (\tilde{\pi}^0\tilde{\pi}^0
 + \tilde{\pi}^+\tilde{\pi}^-) + g_2^2\, (c_W Z_{\mu} + s_W A_{\mu})^2
 \cdot \tilde{\pi}^+\tilde{\pi}^- +\,...\,,  \label{L-piV}
\end{eqnarray}
where $\tilde{\pi}_{,\mu} \equiv \partial_{\mu}\tilde{\pi}$ notation is
used for brevity. Other
parts of the Lagrangian of the vector-like Technicolor model are not
needed for present purposes and can be found in
Refs.~\cite{VLTC,VLTC-DM}.

It is worth to stress here that in distinction to extended TC
scenarios, in the vector-like TC model the technipion interacts only
with SM gauge bosons $Z,\gamma$ and $W^\pm$ and with constituent
$SU(3)_c$-singlet techniquarks. In practice, this makes the
technipions rather difficult to produce and observe even in rather
light $\sim 100$ GeV mass range.

\section{Technipion production and decay: gauge boson channels}

\begin{figure*}[!h]
\begin{minipage}{0.8\textwidth}
 \centerline{\includegraphics[width=1.0\textwidth]{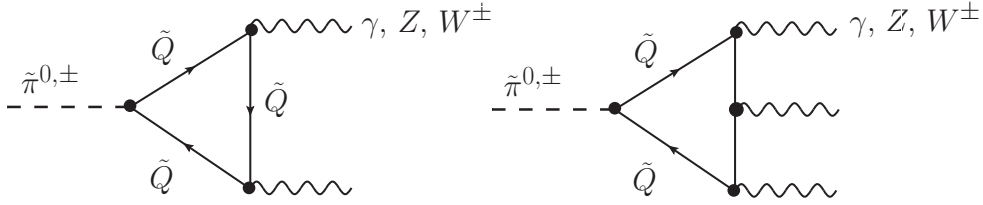}}
\end{minipage}
   \caption{
\small The loop-induced light technipion couplings to the gauge
bosons through constituent techniquark loops. In the case of
$Y_Q\not=0$, the technipion is coupled to two gauge bosons to the
lowest order $\tilde \pi V_1 V_2$ via techniquark triangle diagrams
(left), while for the $Y_Q=0$ case the
technipion is coupled only to three gauge bosons $\tilde \pi V_1 V_2
V_3$ via a box diagram (right). The latter case is much more
involved and will not be considered here.}
 \label{fig:pi-decay}
\end{figure*}

As it follows from Eq.~(\ref{L-piV}), the pseudoscalar technipions
can only be produced in pairs in gauge boson fusion reactions at
Born level while single pion production is possible at one loop
level only. For non-zeroth techniquark hypercharge $Y_Q\not=0$, the
effective one-loop technipion-vector bosons
$\tilde{\pi}^0\,V_1\,V_2$ couplings are given by triangle diagrams
shown in Fig.~\ref{fig:pi-decay} (left). The latter is valid for the
QCD-like TC scenario with $SU(3)_{\rm TC}$ group of confinement
which is the subject of our analysis here.
The corresponding loop amplitude has the following form
\begin{eqnarray}
 && i{\cal V}_{\tilde{\pi}^0\,V_1\,V_2} =
 F_{V_1V_2}(M_1^2,M_2^2,m_{\tilde \pi}^2;M_Q^2)
 \cdot\epsilon_{\mu\nu\rho\sigma}p_1^{\mu}p_2^{\nu}
 {\varepsilon^*_1}^{\rho}{\varepsilon^*_2}^{\sigma}\,,\\
 && F_{V_1V_2}=\frac{N_{\rm TC}}{2\pi^2}\sum_{Q=U,D} g_{V_1}^Q\,
 g_{V_2}^Q\,g_{\tilde{\pi}^0}^Q\,M_{Q}\,
 C_0(M_1^2,M_2^2,m_{\tilde \pi}^2;M_Q^2)\,, \label{FV1V2}
\end{eqnarray}
where $C_0(m_1^2,m_2^2,m_3^2;m^2)\equiv
C_0(m_1^2,m_2^2,m_3^2;m^2,m^2,m^2)$ is the standard finite
three-point function, $N_{\rm TC}$ is the number of technicolors in
confined $SU(N_{\rm TC})$ group, $p_{1,2}$, $\varepsilon_{1,2}$ and
$M_{1,2}$ are the 4-momenta, polarization vectors of the vector
bosons $V_{1,2}$ and their on-shell masses, respectively, and
neutral technipion couplings to $U,D$ techniquarks are
\begin{eqnarray}
g_{\tilde{\pi}^0}^U=g_{\rm TC}\,,\quad g_{\tilde{\pi}^0}^D=-g_{\rm
TC}\,,
\end{eqnarray}
while gauge couplings of techniquarks $g_{V_{1,2}}^Q$ are defined in
Eq.~(\ref{Q-gauge}). We have assumed $m_U = m_D = m_Q$.
We should notice here that the $\tilde \pi^0\to
WW$ decay mode is forbidden by symmetry \cite{VLTC}. Finally, the
explicit expressions of the effective neutral technipion couplings
$F_{V_1V_2}$ for on-shell $V_1V_2=\gamma\gamma$, $\gamma Z$ and $ZZ$
final states are
\begin{eqnarray}
&&F_{\gamma\gamma}=\frac{4\alpha\,g_{\rm
TC}}{\pi}\frac{M_Q}{m_{\tilde
\pi}^2}\,\arcsin^2\Bigl(\frac{m_{\tilde \pi}}{2M_Q}\Bigr)\,, \qquad
\frac{m_{\tilde \pi}}{2M_Q}<1\,,\label{gamgam-pi}\\
&&F_{\gamma Z}=\frac{4\alpha\,g_{\rm TC}}{\pi}\frac{M_Q}{m_{\tilde
\pi}^2}\,\cot2\theta_W\,\Big[\arcsin^2\Bigl(\frac{m_{\tilde
\pi}}{2M_{\tilde Q}}\Bigr)-\arcsin^2\Bigl(\frac{M_Z}{2M_{\tilde
Q}}\Bigr)\Big]\,,\\
&&F_{ZZ}=\frac{2\alpha\,g_{\rm
TC}}{\pi}M_Q\,C_0(M_Z^2,M_Z^2,m_{\tilde \pi}^2;M_{\tilde Q}^2)\,,
\end{eqnarray}
where $\alpha=e^2/4\pi$ is the fine structure constant.

Now the two-body technipion decay width in a vector boson channel
can be represented in terms of the effective couplings (\ref{FV1V2})
as follows:
\begin{eqnarray}
\Gamma(\tilde{\pi}^0\to V_1\,V_2)=r_V\frac{m_{\tilde
\pi}^3}{64\pi}\,\bar{\lambda}^3(M_1^2,M_2^2; m_{\tilde
\pi}^2)\,|F_{V_1V_2}|^2\,,
\end{eqnarray}
where $r_V=1$ for identical bosons $V_1$ and $V_2$ and $r_V=2$ for
different ones, and $\bar{\lambda}$ is the normalized K\"allen
function
\begin{eqnarray}
\bar{\lambda}(m_a,m_b;q)=\Big(1-2\frac{m_a^2+m_b^2}{q^2}+\frac{(m_a^2-m_b^2)^2}{q^4}\Big)^{1/2}\,.
\label{Kallen}
\end{eqnarray}

In Fig.~\ref{fig:pi-decay} (right) we show the leading-order
contribution to single technipion-gauge bosons coupling for $Y_Q=0$
(relevant in the case of an even $SU(N_{\rm TC})_{\rm TC}$ group of
confinement, e.g. $SU(2)_{\rm TC}$ \cite{VLTC-DM}). In the latter
case, a single technipion can be produced in $V_1V_2$ fusion only in
association with an extra gauge boson $V_3$ while produced
technipion should further decay either into three gauge bosons
$\tilde \pi\to V'_1V'_2V'_3$ or into a pair of Higgs bosons $\tilde
\pi\to hh$. Such processes would be rather suppressed and difficult
to study experimentally while they give rise to the only observable
signatures of technipions in the case of $SU(2)_{\rm TC}$ group of
confinement in the vector-like Technicolor scenario so will be
studied elsewhere.

\section{Inclusive technipion production: the VBF mechanism}

\begin{figure*}[!h]
\begin{minipage}{0.6\textwidth}
 \centerline{\includegraphics[width=1.0\textwidth]{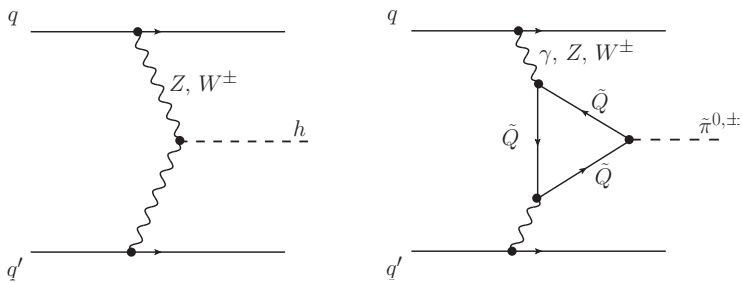}}
\end{minipage}
   \caption{
\small Typical VBF production channels of the Higgs boson at tree
level (left) and technipion via a triangle techniquark loop
(right) via a gauge boson fusion in the quark-(anti)quark
scattering.}
 \label{fig:PiH-prod-diag}
\end{figure*}

Since technipions do not couple directly to SM fermions and gluons,
the only way to produce them is in the vector-boson
($\gamma\gamma,\,\gamma Z,\,ZZ$) fusion channel.
The VBF is typically considered as one of the key
production modes of the Higgs boson at the LHC which allowed recently
for a clear discrimination of the Higgs signal  \cite{ATLAS,CMS}.
Corresponding typical partonic $2\to3$ hard subprocesses
of Higgs boson and $\tilde{\pi}$ production in high
energy hadron-hadron collisions via intermediate VBF mechanism are
shown in Fig.~\ref{fig:PiH-prod-diag}. While VBF
Higgs studies were properly done elsewhere \cite{VLTC}, here we focus on
the VBF into a neutral technipion only.

\begin{figure*}[!h]
\begin{minipage}{0.9\textwidth}
 \centerline{\includegraphics[width=1.0\textwidth]{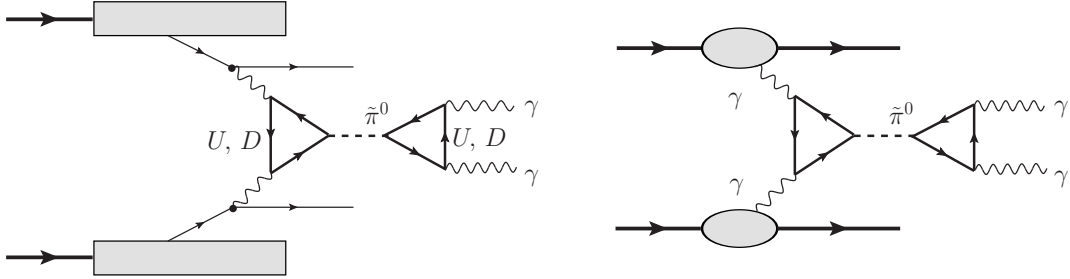}}
\end{minipage}
   \caption{
\small Hadron-level technipion production channels in VBF mechanism
and the leading $\gamma\gamma$ decay channel: inclusive
$\tilde\pi^{0,\pm}$ production in association with two quark jets
(left) and the central exclusive $\tilde\pi^0$ production in the
$\gamma\gamma$ fusion (right).}
 \label{fig:Tpion}
\end{figure*}

In Fig.~\ref{fig:Tpion} we show characteristic diagrams for
the inclusive (left) and central exclusive (right) technipion
production processes in dominant
$\gamma\gamma$ fusion and decay channel. Both, production and decay
subprocesses are initiated by triangle loop of $U,D$ techniquarks.
%
%
We assume $m_U = m_D$.
Thus, the leading-order hard (parton level) VBF subprocess in the
inclusive $h$ (left) and $\tilde \pi$ (right) production in the high
energy $pp$ scattering is quark-initiated one
 \begin{eqnarray}
   q_iq'_j \to q_iq'_j(\gamma^*\gamma^*\to h,\,\tilde \pi^0)\,,
   \label{VBF-pi}
 \end{eqnarray}
where $q_i$ and $q_j$ can be either a quark or an antiquark of
various flavors from each of the colliding protons, and the virtual
$\gamma\gamma$ fusion is concerned. So, the both VBF processes, the
$h$ and $\tilde \pi^0$ production may ``compete''. While $h\to
\gamma\gamma$ branching ratio is very small $\sim 10^{-3}$, the
corresponding $\tilde\pi^0 \to \gamma\gamma$ one is fairly large $\sim 1$.
On the other hand, the Higgs boson has additional dominating
production modes e.g. via gluon-gluon fusion mechanism and the
Higgsstrahlung off gauge bosons and heavy flavor. In contrast to the
Higgs boson production, one technipion can be produced only via heavy
techniquark triangle loop in the VBF mechanism. Such observable
signatures similar to those of the Higgs boson open an
interesting and straightforward opportunity for technipion searches
in standard Higgs boson studies at the LHC.

The calculation of the inclusive production cross sections in QCD is
rather straightforward and based upon standard collinear
factorisation technique so we do not discuss it here. In numerical
estimations of these cross sections which will be discussed later in
the Results section, it is naturally assumed that the incoming quark
$q_i$ and (anti)quark $q'_j$ loose only a small fraction of their
initial energy taken away by intermediate vector bosons. In this
kinematics, the final-state quarks are seen as forward-backward hard
jets, and by measuring their momenta one accurately reconstructs the
invariant mass of the produced state. As was advocated in
Ref.~\cite{VLTC}, an overall one-technipion production rate is
strongly suppressed compared to the Higgs boson production rate,
which along with extremely narrow technipion resonance makes it
rather hard to study experimentally. So, even light technipions down
to $W$ boson mass may be not excluded yet by LEP II and LHC studies,
and the latter point is an interesting subject for further investigations.

\section{Exclusive technipion production: the VBF mechanism}

Now we consider the central exclusive $p p \to p p \tilde{\pi}^0$
process illustrated in Fig.~\ref{fig:Tpion} (right). Similarly to the
inclusive case discussed above, this process is determined by the
colorless VBF subprocess. We take into account only for dominating
$\gamma \gamma \to \tilde{\pi}^0$ fusion reaction and omit $\gamma Z
\to \tilde{\pi}^0$, $Z \gamma \to \tilde{\pi}^0$ and $Z Z \to
\tilde{\pi}^0$ subprocesses which turn out to be numerically very
small being suppressed by large masses in propagators. The
corresponding matrix element for the hadron-level $2 \to 3$ process
can be written as:
\begin{eqnarray}
{\cal M}_{\lambda_a \lambda_b \to \lambda_1 \lambda_2}^{p p \to p p
\tilde\pi^0} &=& V^{\mu_1}_{\lambda_a \to \lambda_1} \frac{(-i
g_{\mu_1 \nu_1})}{t_1} F_{\gamma \gamma}(M_Q,m_{\tilde\pi})
\epsilon^{\nu_1 \nu_2 \alpha \beta} q_{1, \alpha} q_{2, \beta}
\frac{(-i g_{\mu_2 \nu_2})}{t_2} V^{\mu_2}_{\lambda_b \to
\lambda_2}\,,
\end{eqnarray}
where the parton-level triangle amplitude
$F_{\gamma\gamma}(M_Q,m_{\tilde\pi})$ is given by
Eq.~(\ref{gamgam-pi}), and the vertex functions $V_{\mu_{1,2}}$ can
be approximated in the spin conserving case relevant at high
energies as follows
\begin{eqnarray}
V^{\mu_1}_{\lambda_a \to \lambda_1} \simeq F_1(t_1) \bar
u(\lambda_1) i \gamma^{\mu_1} u(\lambda_a)\,, \quad
V^{\mu_2}_{\lambda_b \to \lambda_2} \simeq F_1(t_2) \bar
u(\lambda_2) i \gamma^{\mu_2} u(\lambda_b)\,,
\end{eqnarray}
where $F_1(t)$ is the electromagnetic proton form factor. The
natural limitation for a light pseudo-Goldstone technipion
\begin{equation}
\frac{m_{\tilde\pi}}{2 M_Q} < 1
\end{equation}
is implied. The matrix element specified above is used in a
three-body calculation precisely as for the usual exclusive pion
production in the $p p \to p p \pi^0$ process considered in
Ref.~\cite{LS13}.

\section{Exclusive $\gamma\gamma$ background: QCD vs QED mechanisms}

\begin{figure*}[!h]
\begin{minipage}{0.8\textwidth}
 \centerline{\includegraphics[width=1.0\textwidth]{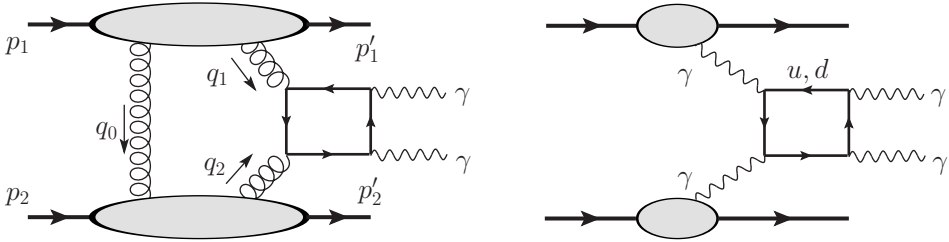}}
\end{minipage}
   \caption{
\small Irreducible non-resonant background processes for the central
exclusive technipion $\tilde\pi^0\to\gamma\gamma$ production in $pp$
collisions at the LHC: the QCD diffractive $\gamma\gamma$ pair
production (left) and the QED-initiated $\gamma\gamma$ pair
production (right). In the latter case, only a part of contributions
corresponding to quark boxes is shown here for illustration while in
actual calculations the full set of SM contributions including quark,
lepton and $W$ boson loops is taken into account.}
 \label{fig:gamgam-CEP}
\end{figure*}

In order to estimate the feasibility of exclusive technipion
production studies we need to analyze carefully the exclusive
$\gamma\gamma$ background. There are two basic non-resonant leading
order box-induced contributions -- the QCD (Durham) diffractive
mechanism via $gg\to \gamma\gamma$ shown in
Fig.~\ref{fig:gamgam-CEP} (left) and the QED (light-by-light)
scattering mechanism $\gamma\gamma\to\gamma\gamma$ shown in
Fig.~\ref{fig:gamgam-CEP} (right). Below, we discuss both of them in
detail.

\subsection{Durham QCD mechanism}

A schematic diagram for central exclusive production of
$\gamma\gamma$ pairs in proton-proton scattering $pp\to
p(\gamma\gamma)p$ with relevant kinematics notations is shown in
Fig.~\ref{fig:gamgam-CEP} (left). In what follows, we use the
standard theoretical description of CEP processes developed by the
Durham group for the exclusive production of Higgs boson in
Ref.~\cite{Durham}. The details of the kinematics for the central
exclusive production processes can be found e.g. in
Ref.~\cite{Albrow:2010yb}. Here we only sketch basic notations used
in our calculations, which are similar to those in our previous
paper on the central exclusive production of $W^+W^-$ pairs
\cite{LPS13}.

The momenta of intermediate gluons are given by Sudakov
decomposition in terms of the incoming proton four-momenta
$p_{1,2}$
\begin{eqnarray}\nonumber
&&q_1=x_1p_1+q_{1\perp},\quad q_2=x_2p_2+q_{2\perp},\quad 0<x_{1,2}<1,\\
&&q_0=x'p_1-x'p_2+q_{0\perp}\simeq q_{0\perp},\quad x'\ll x_{1,2},
\label{moms}
\end{eqnarray}
where $x_{1,2},x'$ are the longitudinal momentum fractions for
active (fusing) and color screening gluons, respectively, such that
$q_{\perp}^2\simeq -|\bf q_{\perp}|^2$.

The QCD factorisation of the process at the hard scale $\mu_F$ is
provided by the large invariant mass of the $\gamma\gamma$ pair
$M_{\gamma\gamma}$, i.e.
\begin{eqnarray}\label{sx1x2}
\mu_F^2\equiv s\,x_1x_2\simeq M_{\gamma\gamma}^2\,.
\end{eqnarray}
It is convenient to introduce the Sudakov expansion for photon
momenta as follows
\begin{eqnarray}
k_3=x_1^+ p_1+x_2^+ p_2+k_{3\perp},\quad
k_4=x_1^- p_1+x_2^- p_2+k_{4\perp}
\end{eqnarray}
leading to
\begin{eqnarray}\label{xqq}
x_{1,2}=x_{1,2}^+ + x_{1,2}^-,\quad
x_{1,2}^+=\frac{|{\bk}_{3,4\perp}|}{\sqrt{s}}e^{\pm y_3},\quad
x_{1,2}^-=\frac{|{\bk}_{3,4\perp}|}{\sqrt{s}}e^{\pm y_4}
\end{eqnarray}
in terms of photon rapidities $y_{\pm}$ and transverse masses
$m_{3,4\perp}$. For simplicity, in actual calculations we work in
the forward limit which implies that ${\bk}_{3\perp}\simeq
-{\bk}_{4\perp}$.

We write the amplitude of the diffractive process, which at high
energy is dominated by its imaginary part, as
\begin{eqnarray} \label{ampl}
{\cal M}_{\lambda_3\lambda_4}(s,t_1,t_2) &\simeq&is\frac{\pi^2}{2}
\int d^2 {\bq}_{0\perp} V_{\lambda_3\lambda_4}(q_1,q_2,k_{3},k_{4})
\frac{f_g(q_0,q_1;t_1)f_g(q_0,q_2;t_2)}
{{\bq}_{0\perp}^2\,{\bq}_{1\perp}^2\,{\bq}_{2\perp}^2}\,,
\end{eqnarray}
where $\lambda_{3,4}=\pm 1,\,0$ are the polarisation states of the
produced photons, respectively, $f_g(r_1,r_2;t)$ is the off-diagonal
unintegrated gluon distribution function (UGDF), which depends on
the longitudinal and transverse components of both gluon momenta.
The gauge-invariant $gg\to \gamma_{\lambda_3} \gamma_{\lambda_4}$
hard subprocess amplitude
$V_{\lambda_3\lambda_4}(q_1,q_2,k_{3},k_{4})$ is given by
the light cone projection
\begin{eqnarray}\label{GIproj}
V_{\lambda_3\lambda_4}=
n^+_{\mu}n^-_{\nu}V_{\lambda_3\lambda_4}^{\mu\nu}= \frac{4}{s}
\frac{q^{\nu}_{1\perp}}{x_1} \frac{q^{\mu}_{2\perp}}{x_2}
V_{\lambda_3\lambda_4,\mu\nu},\quad
q_1^{\nu}V_{\lambda_3\lambda_4,\mu\nu}=
q_2^{\mu}V_{\lambda_3\lambda_4,\mu\nu}=0\,,
\end{eqnarray}
where $n_{\mu}^{\pm} = p_{1,2}^{\mu}/E_{p,cms}$ and the
center-of-mass proton energy $E_{p,cms} = \sqrt{s}/2$. We adopt the
definition of gluon polarisation vectors proportional to transverse
momenta $q_{1,2 \perp}$, i.e. $\epsilon_{1,2} \sim q_{1,2 \perp} /
x_{1,2}$. The helicity matrix element in the previous expression
reads
\begin{eqnarray}
V_{\lambda_3\lambda_4}^{\mu\nu}(q_1,q_2,k_{3},k_{4})
=\varepsilon^{\rho}(k_3,\lambda_3) \varepsilon^{\sigma}(k_4,\lambda_4)
V_{\rho\sigma}^{\mu\nu}\,, \label{Vepsilon}
\end{eqnarray}
in terms of the Lorentz and gauge invariant $2\to2$ amplitude
$V_{\rho\sigma}^{\mu\nu}$ and photons polarisation vectors
$\varepsilon(k,\lambda)$. Below we will analyze the exclusive
production with polarized photons. In Eq.~(\ref{Vepsilon})
$\varepsilon_{\mu}(k_3,\lambda_3)$ and $\varepsilon_{\nu}(k_4,\lambda_4)$
can be defined easily in the proton-proton center-of-mass frame with
$z$-axis along the proton beam as
\begin{eqnarray}
\varepsilon(k,\pm 1) &=& \frac{1}{\sqrt{2}} \left(0,\, i\sin\phi \mp
\cos\theta\cos\phi,\, -i\cos\phi \mp \cos\theta\sin\phi,\, \pm
\sin\theta\right)\,,
\label{vectors}
\end{eqnarray}
where $\phi$ is the azimuthal angle of a produced photon, and
$\varepsilon^{\mu}(\lambda)\varepsilon^*_{\mu}(\lambda)=-1$ and
$\varepsilon_{\mu}(k_3,\lambda_3)k_3^{\mu}=
 \varepsilon_{\nu}(k_4,\lambda_4)k_4^{\nu}=0$.
In the forward scattering limit, the azimuthal angles of the final
state photons are related as $\phi_3 = \phi_4 + \pi$.

The diffractive amplitude given by Eq.~(\ref{ampl}) is averaged over
the color indices and over the two transverse polarizations of the
incoming gluons. The relevant color factor which includes summing
over colors of quarks in the box loop and averaging over fusing
gluon colors (according to the definition of unintegrated gluon
distribution function) is the same as in the previously studied
Higgs CEP \cite{MPS_bbbar} (for more details on derivation of
the generic $pp\to pXp$
amplitude, see e.g. Ref.~\cite{Albrow:2010yb}). The matrix element
$V_{\lambda_{3},\lambda_{4}}$ contains twice the strong coupling
constant $g_s^2 = 4 \pi \alpha_{s}$. In our calculation here we take
the running coupling constant $\alpha_s(\mu_{\rm
hard}^2=M_{\gamma\gamma}^2)$ which depends on the invariant mass of
$\gamma\gamma$ pair as a hard renormalisation scale of the process.
The choice of the scale introduces roughly a factor of
two uncertainty when varying the hard scale $\mu_{\rm
hard}$ between $2M_{\gamma\gamma}$ and $M_{\gamma\gamma}/2$.

The bare amplitude above is subjected to absorption corrections that
depend on the collision energy and typical proton transverse
momenta. As was done in original Durham calculations \cite{Durham},
the bare production cross section is usually multiplied by a gap
survival factor which we take the same as for the Higgs boson and $b
\bar b$ production to be $S_{g} = 0.03$ at the LHC energy (see e.g.
Ref.~\cite{MPS2011_gg}).

The diffractive $\gamma \gamma$ CEP amplitude (\ref{ampl}) described
above is used now to calculate the corresponding cross section
including realistic limitations on the phase space. For the sake of
simplicity, assuming an exponential slope of $t_{1,2}$-dependence of
the UGDFs \cite{Durham}, and as a consequence of the approximately
exponential dependence of the cross section on $t_1$ and $t_2$
(proportional to $\exp(b t_1)$ and $\exp(b t_2)$), the four-body
phase space can be calculated as
\begin{eqnarray}
d \sigma \approx && \frac{1}{2s}\overline{ |{\cal
M}|^2}\Big|_{t_{1,2}=0}\,\frac{1}{2^4}
\frac{1}{(2\pi)^8}\frac{1}{E'_1E'_2}\, \frac14\,
\frac{1}{b^2}\,(2\pi)^2\, \frac{p_{m\perp}}{4}{\cal J}^{-1}\,dy_3
dy_4 dp_{m\perp} d\phi_m \,.
\label{redPS}
\end{eqnarray}
Since in this approximation we have assumed no correlations between
outgoing protons (which is expected here and is practically true for
the production of $b \bar b$ \cite{MPS_bbbar} or $g g$
\cite{MPS2011_gg} dijets) there is no dependence of the integrand in
Eq.~(\ref{redPS}) on $\phi_m$, which means that the phase space
integration can be further reduced to three-dimensional one. The
Jacobian ${\cal J}$ in Eq.~(\ref{redPS}) is given by \cite{LS2010}
\begin{eqnarray}
{\cal J}=\Bigg| \frac{p_{1z}'}{\sqrt{m_p^2+{p'}_{1z}^2}} -
\frac{p_{2z}'}{\sqrt{m_p^2+{p'}_{2z}^2}} \Bigg|   \; .
\end{eqnarray}
In actual calculations below we shall use the reduced form of the
four-body phase space Eq.~(\ref{redPS}), and it is checked to give
correct numerical results against the full phase space calculation
for some simple reactions. Different representations of the phase
space depending on a particular kinematical distributions needed can
be found in Ref.~\cite{LS2010}.

\begin{figure*}[!h]
\begin{minipage}{0.9\textwidth}
 \centerline{\includegraphics[width=1.0\textwidth]{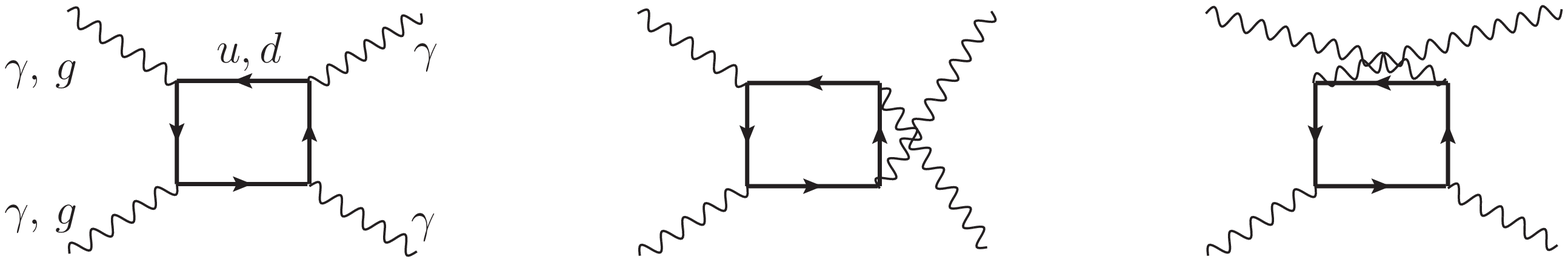}}
\end{minipage}
   \caption{
\small Representative topologies of the hard subprocesses $gg\to
\gamma\gamma$ and $\gamma\gamma \to \gamma\gamma$, which contribute
to exclusive $\gamma\gamma$ pair production. These subprocesses
constitute the irreducible background for the exclusive
$\tilde{\pi}^0\to\gamma\gamma$ reaction at the LHC. In the $gg \to
\gamma\gamma$ case only quarks propagate in boxes and the amplitude
is dominated by light quarks. In the $\gamma\gamma \to \gamma\gamma$
case, all the charged fermions -- quarks, leptons, as well as
$W^\pm$ bosons participate in the corresponding box diagrams. In the
latter case, only a part of contributions corresponding to quark
boxes is shown here for illustration.}
 \label{fig:GGhard}
\end{figure*}

Typical contributions to the leading order $gg\to \gamma\gamma$
subprocess are shown in Fig.~\ref{fig:GGhard}. The total number of
topologically different loop diagrams in the Standard Model amounts
to twelve boxes. So the $\gamma\gamma$ background does not exhibit
resonant features which is good for probing New Physics
$\gamma\gamma$-resonant contributions like the technipion signal
under consideration.

The box contributions to the $gg\to \gamma\gamma$ parton level
subprocess amplitude in Fig.~\ref{fig:GGhard} for on-shell fusing
gluons were calculated analytically by using the Mathematica-based
{\tt FormCalc} (FC) \cite{FC} package. The complete matrix element
was automatically generated by FC tools in terms of one-loop
Passarino-Veltman two-, three- and four-point functions and other
internally-defined functions (e.g. gluon and vector bosons
polarisation vectors) and kinematical variables. In the next step,
the Fortran code for the matrix element was generated, and then used
as an external subroutine in our numerical calculations together
with other FC routines setting up the Standard Model parameters,
coupling constants and kinematics. Instead of built-in FC
polarisation vectors we have used transverse gluon polarisation
vectors which enter the projection in Eq.~(\ref{GIproj}), and the
standard photon polarisation vectors defined in Eq.~(\ref{vectors}),
giving an access to individual polarisation states of the
photons. In accordance with the $k_t$-factorisation technique, the
gauge invariance of the resulting amplitudes for the on-mass-shell
initial gluons is ensured by a projection onto the gluon transverse
polarisation vectors proportional to the transverse gluon momenta
$q_{1,2\perp}$ according to Eq.~(\ref{GIproj}).

For the evaluation of the scalar master tree- and four-point
integrals in the gluon-gluon fusion subprocess we have used the {\tt
LoopTools} library \cite{FC}. The result is summed up over all
possible quark flavors in loops and over distinct loop topologies.
We have also checked that the sum of relevant diagrams is explicitly
finite and obeys correct asymptotical properties and energy
dependence. It is worth to mention that a large cancelation between
separate box contributions in the total sum of diagrams takes place,
which is expected from the general Standard Model symmetry
principles.

As soon as the hard subprocess matrix element (denoted above as
$V_{\lambda_3\lambda_4}$) has been defined as a function of relevant
kinematical variables (four-momenta of incoming/outgoing particles),
the loop integration over $q_{0\perp}$ in Eq.~(\ref{ampl}) was
performed to obtain the diffractive amplitude, which then has been
used to calculate the differential distributions for (un)polarised
photon in an external phase space integrator.

\subsection{QED-initiated $\gamma \gamma \to \gamma\gamma$ reaction}

In this subsection, we briefly discuss the mechanism of exclusive
production of two photons via hard $\gamma \gamma \to \gamma \gamma$
subprocess as illustrated in Fig.~\ref{fig:gamgam-CEP} (left).

The light-by-light $\gamma \gamma \to \gamma \gamma$ scattering
subprocess to the leading and next-to-leading order was discussed
earlier in the literature (see e.g. Refs.~\cite{JT94,Dixon}). The
relevant subprocess diagrams are similar in topology to those for
$gg\to \gamma\gamma$ shown in Fig.~\ref{fig:GGhard} but contain
extra contributions from leptonic and vector boson $W$ loops. The
next-to-leading order corrections \cite{Dixon} were found to be
rather small. So in the present paper with the focus on $p p \to p p
\gamma \gamma$ process we consider the leading-order approximation
for the $\gamma \gamma \to \gamma \gamma$ subprocess only.

The cross section of exclusive $\gamma\gamma$ production in $pp$
scattering can be calculated in the same way as in the parton model
in the so-called equivalent photon approximation as
\begin{equation}
\frac{d \sigma}{d y_3 d y_4 d^2 p_{\gamma\perp}}
= \frac{1}{16 \pi^2 {\hat s}^2}
x_1 \gamma(x_1) x_2 \gamma(x_2)
\overline{|
{\cal M}_{\gamma\gamma \to \gamma\gamma}
(\lambda_1, \lambda_2, \lambda_3,\lambda_4) |^2} \; .
\label{exclusive_gamgam}
\end{equation}
A more involved and precise four-body calculation for the $p p \to p
p \gamma \gamma$ is expected to give a very similar result
\cite{LPS13}.

In the parton formula above, $\gamma(x)$ is an elastic flux
($x$-distribution) of equivalent photons associated with elastic
electromagnetic emission off a proton. In practical calculations we
shall use parametrization proposed in Ref.~\cite{DZ89}. In the same
way as for QCD diffractive mechanism described above, the
loop-induced helicity matrix elements for the $\gamma \gamma \to
\gamma \gamma$ subprocess were calculated by using {\tt LoopTools}
\cite{FC}. In numerical calculations we include box diagrams with
lepton, quark as well as with $W$ bosons. At high diphoton invariant
masses the inclusion of diagrams with $W$ bosons is crucial. In
principle, effects beyond the Standard Model possibly responsible
for anomalous gauge couplings could be important
\cite{JT94,OWZ94,GS98,D99,Ch00}, so the exclusive non-resonant
$\gamma\gamma$ background is very interesting by itself. In the
present analysis we concentrate on the search for technipion so we
ignore effects beyond the Standard Model as far as the background is
considered.

\section{Results}

\begin{figure*}[!h]
\begin{minipage}{0.48\textwidth}
 \centerline{\includegraphics[width=1.0\textwidth]{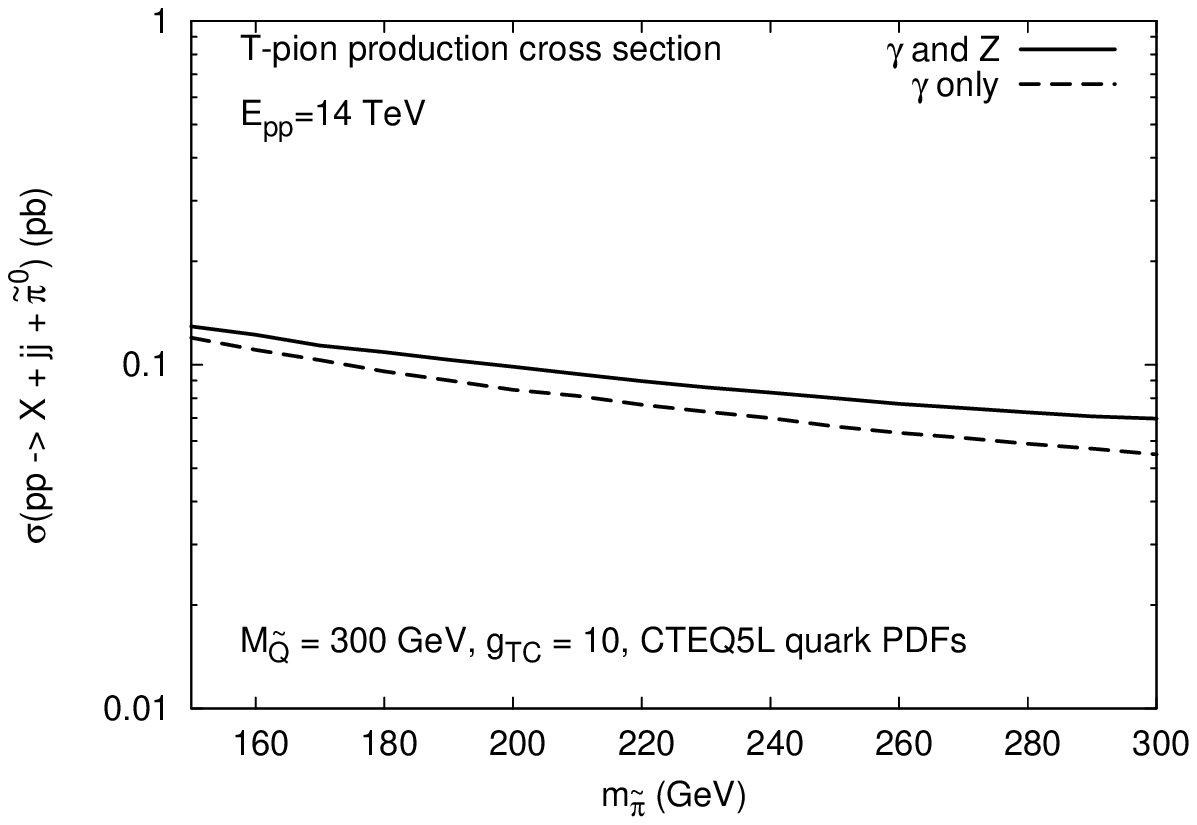}}
\end{minipage}
\begin{minipage}{0.48\textwidth}
 \centerline{\includegraphics[width=1.0\textwidth]{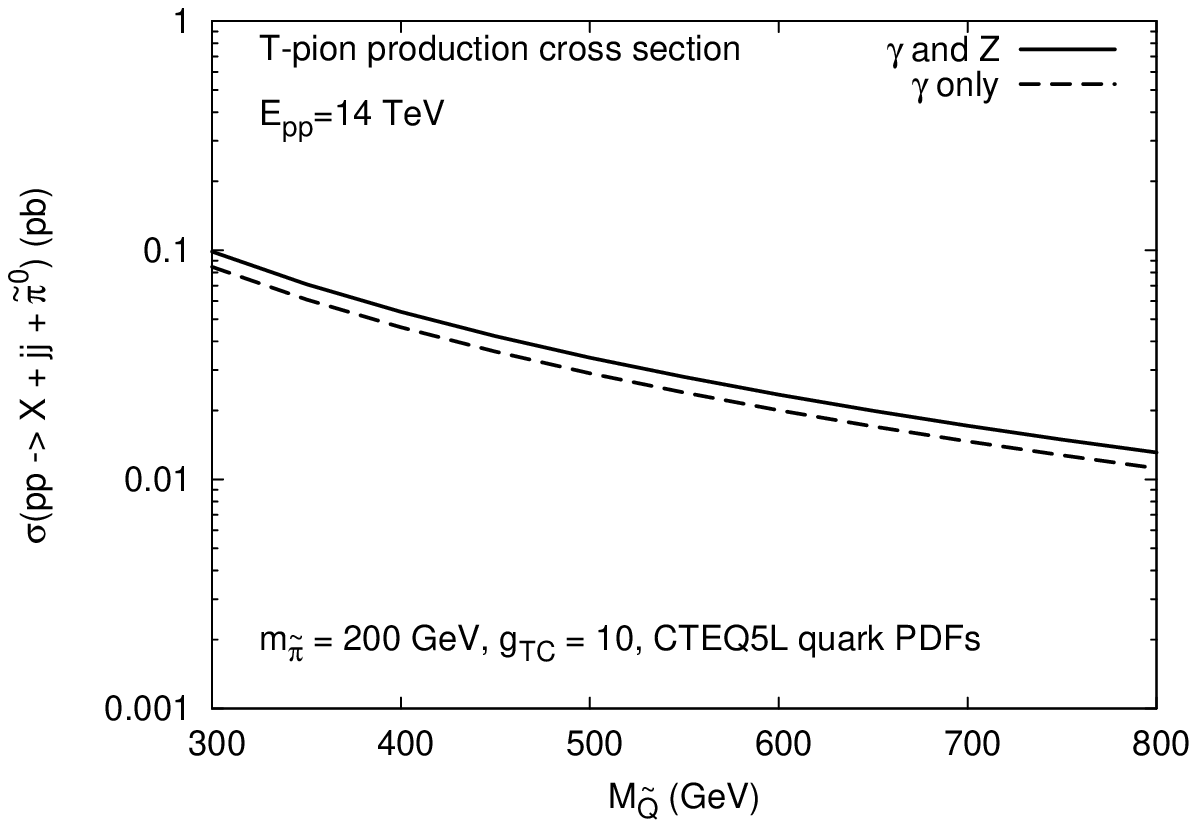}}
\end{minipage}
   \caption{
\small Inclusive $\tilde \pi^0$ production cross section in
association with two forward jets as a function of technipion mass
(left panel) and as a function of techniquark mass (right panel) for
fixed values of the $g_{tc}$ coupling constant at the nominal
LHC energy $\sqrt{s}$ = 14 TeV.}
 \label{fig:inclusive}
\end{figure*}

Before discussing results for exclusive production of neutral
technipion, we would like to summarize the inclusive $\tilde\pi^0$
production in association with two forward jets. In
Fig.~\ref{fig:inclusive} we show the total inclusive cross section
as a function of technipion (left) and techniquark (right) masses,
$m_{\tilde \pi}$ and $M_{\tilde Q}$, respectively, and integrated
over the full phase space.
The calculation was performed in the collinear QCD factorization
with hard (parton-level) 2 $\to$ 3 subprocess (\ref{VBF-pi})
including $t$-channel exchanges of $\gamma$ and $Z^0$ bosons as
illustrated in Fig.~\ref{fig:Tpion} (left) (for more details we
refer to Ref.~\cite{VLTC}). This calculation includes all the light
quark and antiquark flavors in the initial state with respective
quark PDFs. As can be seen from Fig.~\ref{fig:inclusive} the
photon-photon $\gamma\gamma$ fusion mechanism dominates, while
$Z\gamma$ and $ZZ$ fusion contributions are always small (suppressed
by a large mass of $Z$ boson in propagators). The cross section for the
vector-like TC model parameters and CTEQ5L quark PDFs
\cite{Lai:1999wy} chosen as indicated in the figure is of the order
of 100 fb.
\begin{figure*}[!h]
\begin{minipage}{0.5\textwidth}
 \centerline{\includegraphics[width=1.0\textwidth]{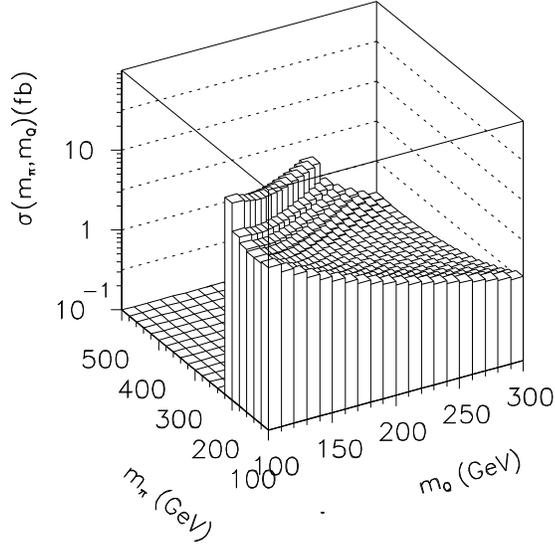}}
\end{minipage}
   \caption{
\small Exclusive cross section as a 2D function of technipion mass
($m_{\tilde \pi}$) and techniquark mass ($M_{\tilde Q}$) for a fixed
value of $g_{\rm TC}$ = 10.}
 \label{fig:exclusive_map}
\end{figure*}

Now let us look into the parameter dependence of the exclusive
production cross section. This calculation is performed in the same
way as the calculation for the exclusive production of usual pion
$\pi^0$ studied recently by two of us in Ref.~\cite{LS13}.
\begin{figure*}[!b]
\begin{minipage}{0.32\textwidth}
 \centerline{\includegraphics[width=1.0\textwidth]{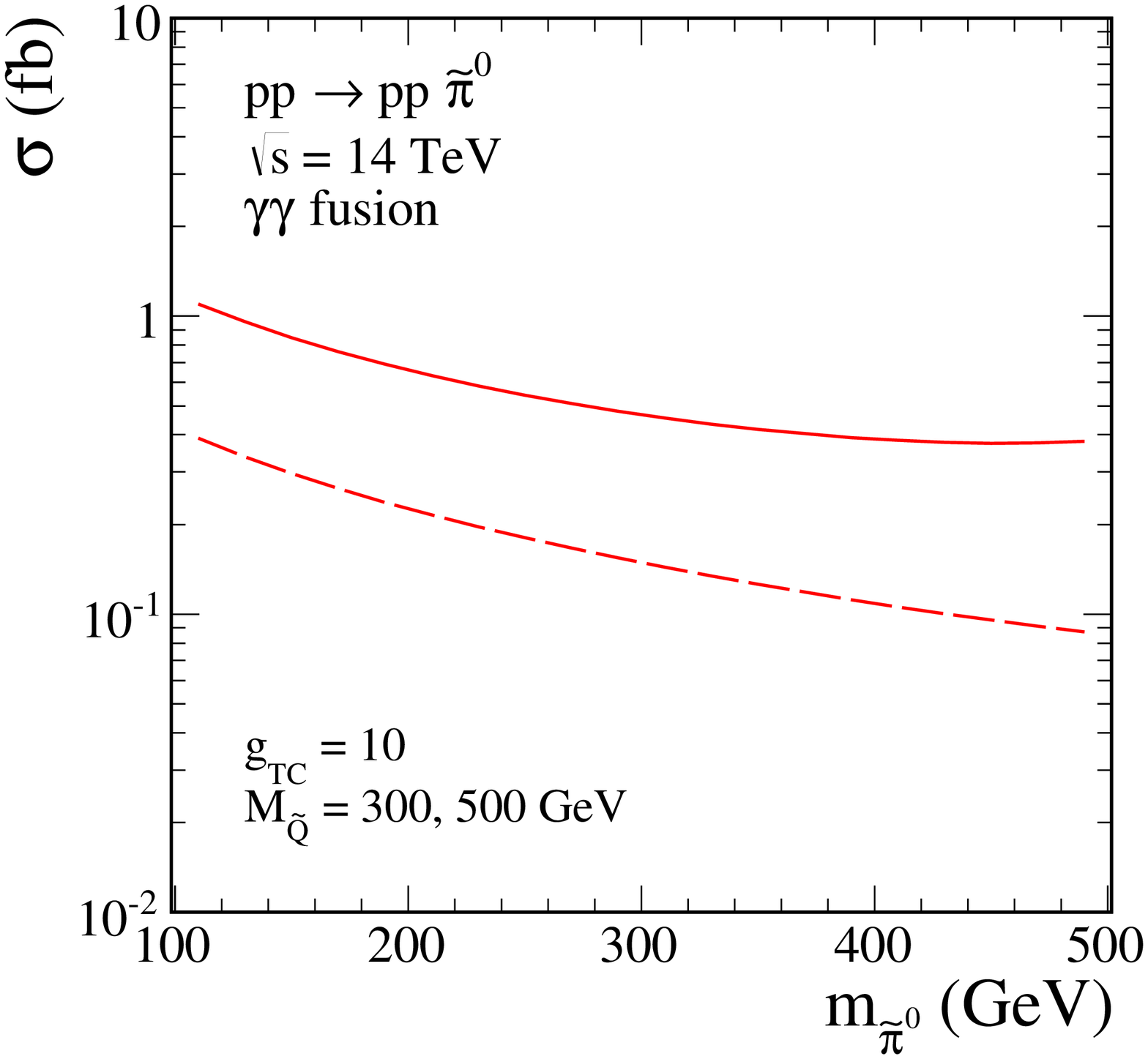}}
\end{minipage}
\begin{minipage}{0.32\textwidth}
 \centerline{\includegraphics[width=1.0\textwidth]{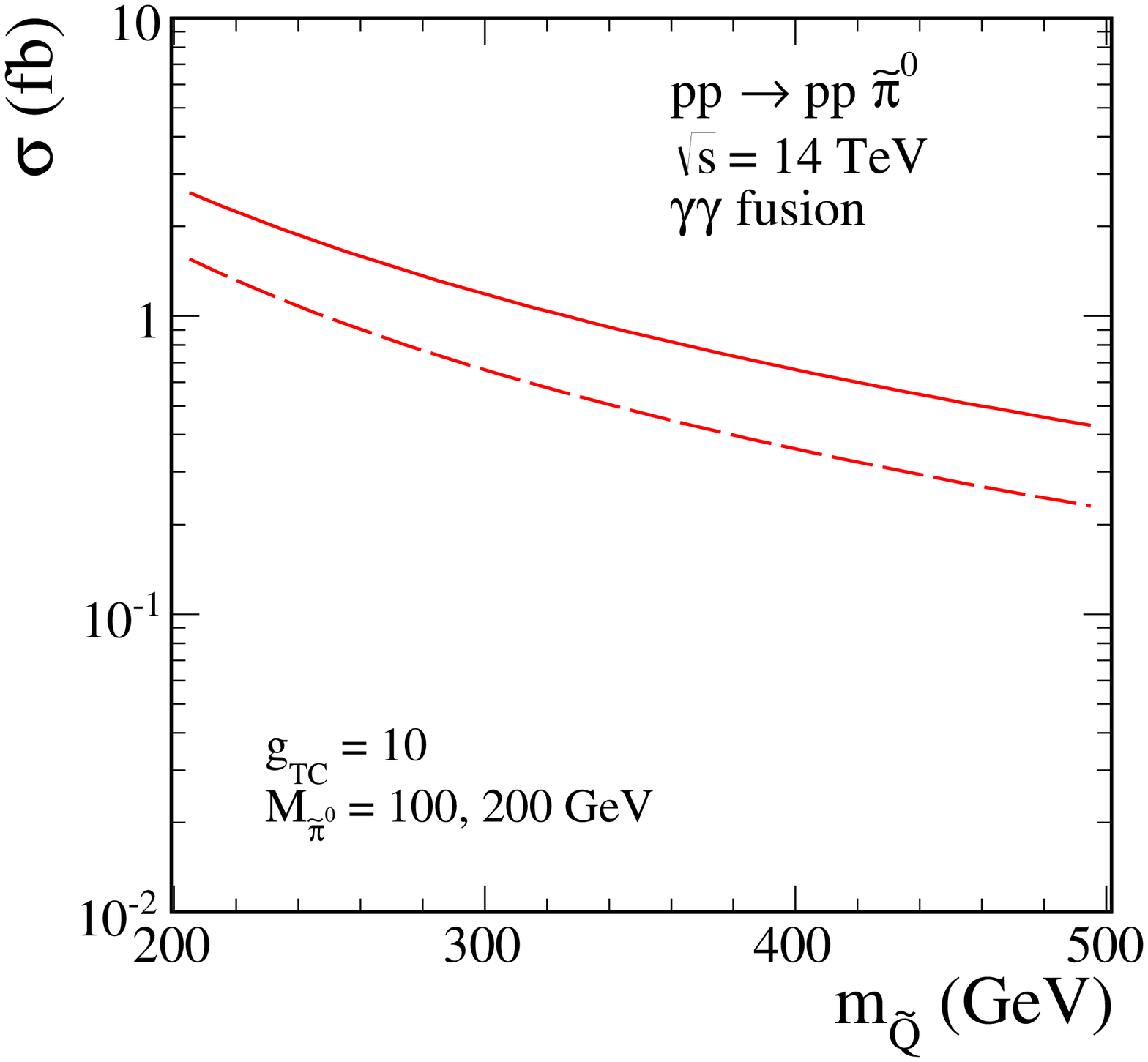}}
\end{minipage}
\begin{minipage}{0.32\textwidth}
 \centerline{\includegraphics[width=1.0\textwidth]{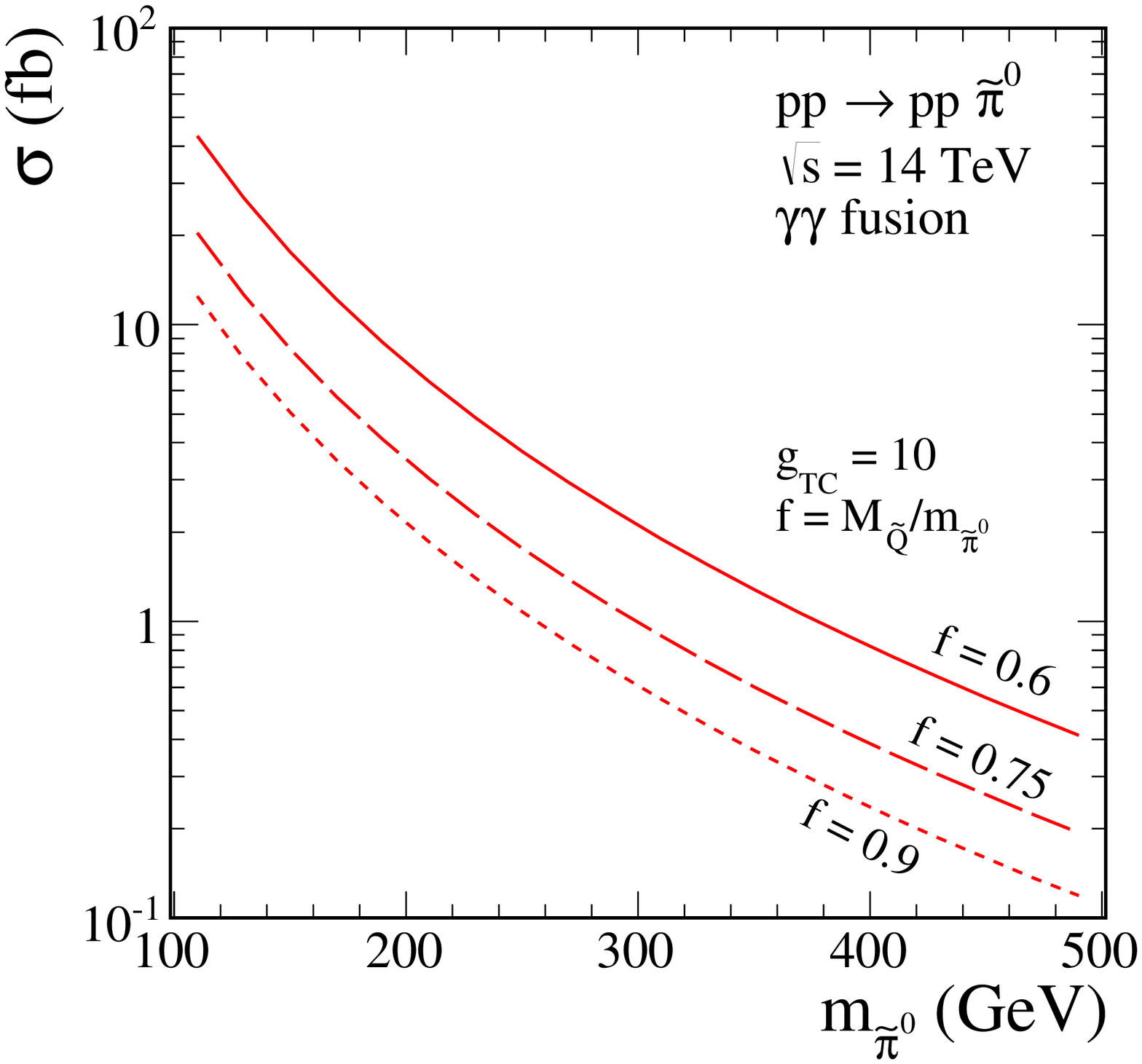}}
\end{minipage}
   \caption{
\small Integrated exclusive cross section as a function of
technipion mass (left) and techniquark mass (middle) for fixed
remaining model parameters as specified in the figure. In the right
panel we show the cross section as a function of technipion mass for
a few fixed ratios $f=M_{\tilde Q} / m_{\tilde \pi}$.}
 \label{fig:parameters}
\end{figure*}
In particular, Fig.~\ref{fig:exclusive_map} shows a 2D map of the full
phase space integrated cross section as a function of technipion and
techniquark masses. A kinematical limit $m_{\tilde \pi} = 2
M_{\tilde Q}$ is clearly visible. We obtain the cross section of the
order of 1 fb for the same parameters as used in the calculation of
the inclusive cross section. This is about two orders of magnitude
less than in the inclusive case. The signal-to-background ratio, as
will be discussed later is, however, more advantageous in the
exclusive case than in the inclusive one.
\begin{figure*}[!h]
\begin{minipage}{0.4\textwidth}
 \centerline{\includegraphics[width=1.0\textwidth]{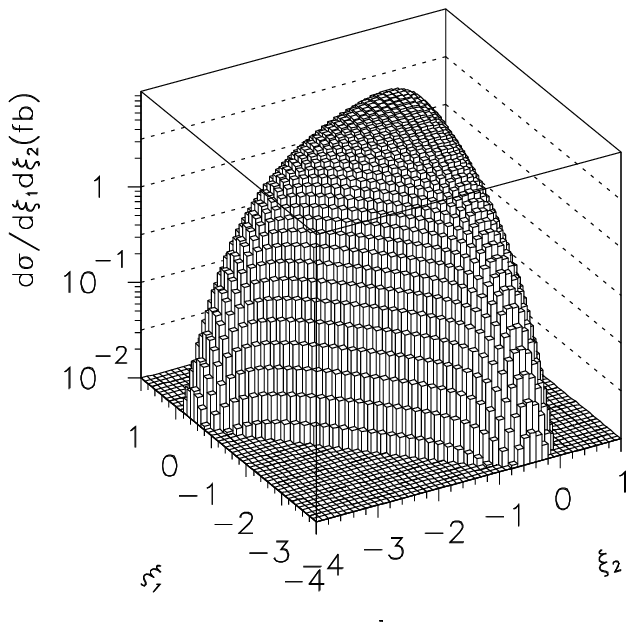}}
\end{minipage}
\begin{minipage}{0.3\textwidth}
 \centerline{\includegraphics[width=1.0\textwidth]{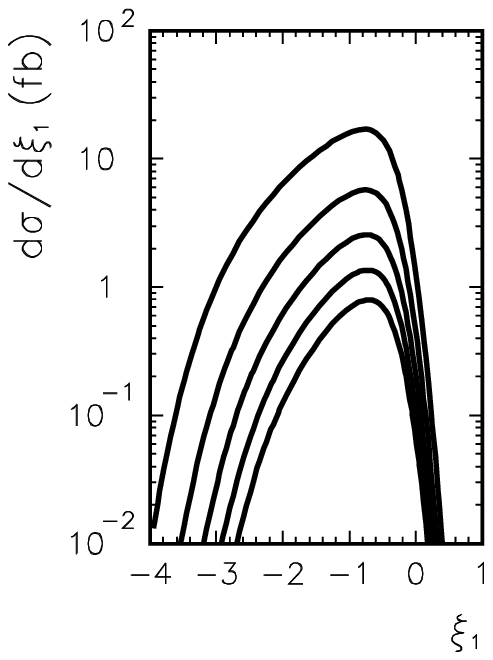}}
\end{minipage}
   \caption{
\small Two-dimensional distribution in the auxiliary quantities
$\xi_1 = log_{10}(p_{1t}/1 \mbox{GeV})$ and
$\xi_2 = log_{10}(p_{2t}/1 \mbox{GeV)}$
(left) and the projection on one of the axes (right).}
 \label{fig:xi1xi2}
\end{figure*}

In Fig.~\ref{fig:parameters} we show one-dimensional dependencies on
technipion (left) and techniquark (middle) masses. These
dependencies can be compared to those in Fig.~\ref{fig:inclusive}.
Finally in Fig.~\ref{fig:parameters} (right) we show dependence on
technipion mass for fixed ratio of techniquark-to-technipion mass
ratio. The latter dependence looks, however, steeper as an artifact
of parameter correlations.
\begin{figure*}[!h]
\begin{minipage}{0.35\textwidth}
 \centerline{\includegraphics[width=1.0\textwidth]{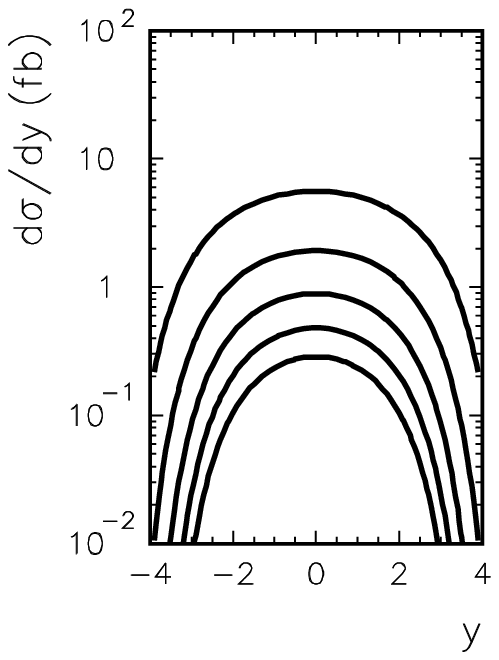}}
\end{minipage}
\begin{minipage}{0.35\textwidth}
 \centerline{\includegraphics[width=1.0\textwidth]{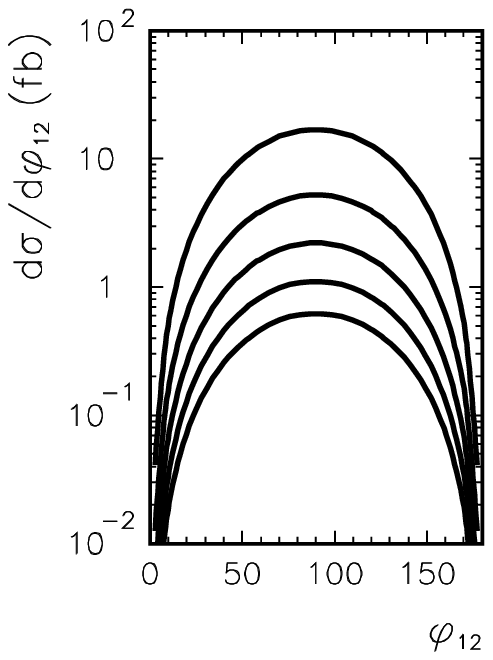}}
\end{minipage}
   \caption{
\small Differential distributions in technipion rapidity (left
panel) and azimuthal angle between outgoing protons (right panel)
for different masses of the technipion ($m_{\tilde \pi}$ = 100, 200,
300, 400, 500 GeV from top to bottom). The techniquark mass is fixed
to be $M_{\tilde Q} = 0.75 m_{\tilde \pi}$.}
 \label{fig:differential_distributions}
\end{figure*}

In the exclusive case, the integration in proton transverse momenta
requires a special care. Instead of integration over $p_{1\perp}$
and $p_{2\perp}$ we integrate over: $\xi_1 = \log_{10}(p_{1\perp}/1
\,{\rm GeV})$ and $\xi_2 = \log_{10}(p_{2\perp}/1\, {\rm GeV})$. The
resulting cross section in the auxiliary quantities is shown in
Fig.~\ref{fig:xi1xi2}.
\begin{figure*}[!h]
\begin{minipage}{0.48\textwidth}
 \centerline{\includegraphics[width=1.0\textwidth]{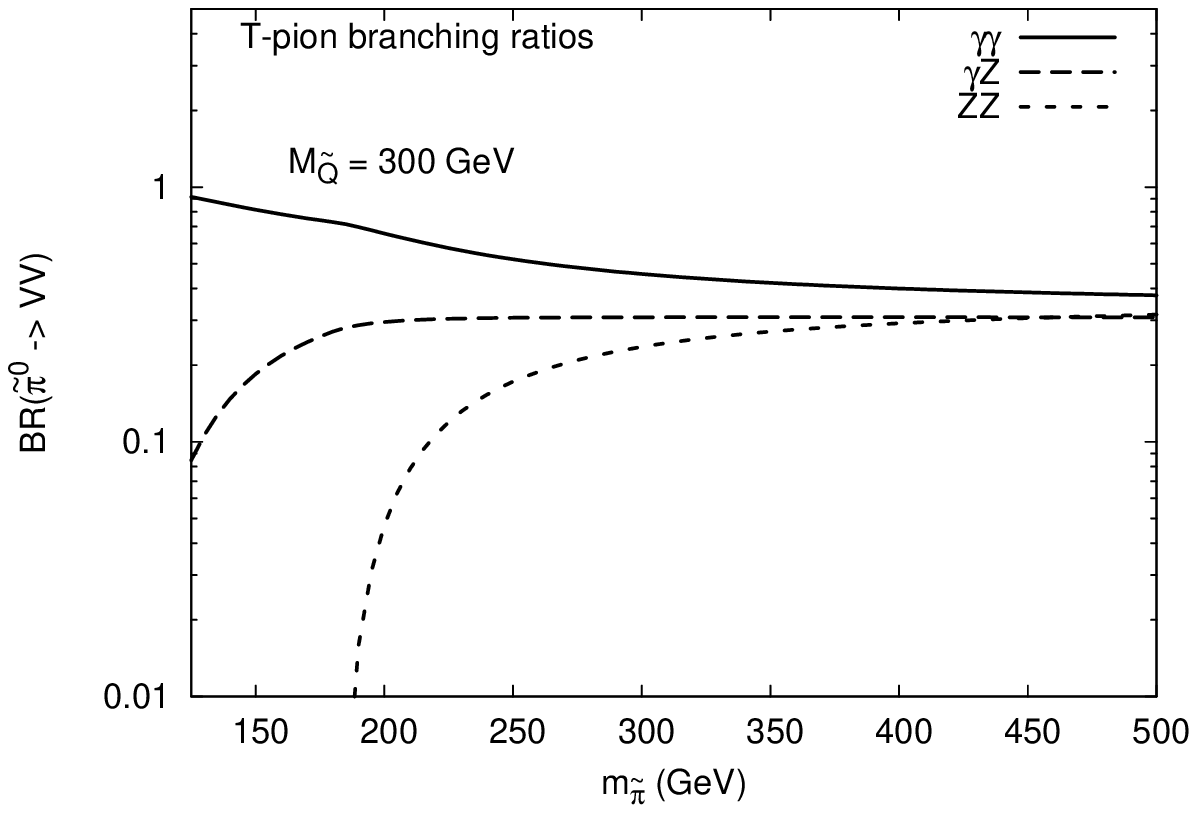}}
\end{minipage}
\begin{minipage}{0.48\textwidth}
 \centerline{\includegraphics[width=1.0\textwidth]{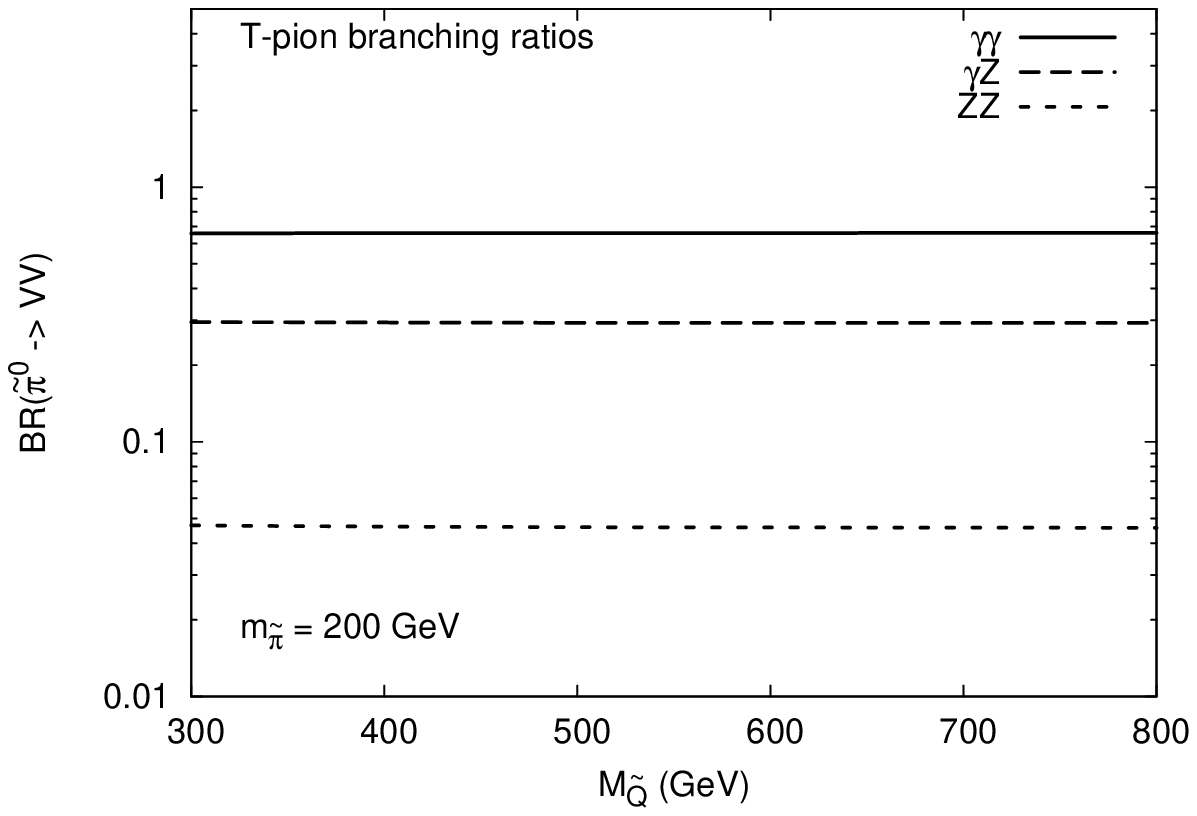}}
\end{minipage}
   \caption{
\small Branching fractions of technipion decays into $\gamma
\gamma$, $\gamma Z$ and $Z Z$ final states as a function of
technipion mass $m_{\tilde \pi}$ for a fixed value of techniquark
mass (left) and as a function of techniquark mass $M_{\tilde Q}$ for
a fixed value of technipion mass (right).}
 \label{fig:branching_fractions}
\end{figure*}

Now let us consider some important differential distributions. In
Fig.~\ref{fig:differential_distributions} we show a distribution in
technipion rapidity (left panel) and azimuthal angle between
outgoing protons (right panel). The larger the technipion mass the
smaller the cross section. The technipions are produced dominantly
at midrapidities as expected.
\begin{figure*}[!h]
\begin{minipage}{0.4\textwidth}
 \centerline{\includegraphics[width=1.0\textwidth]{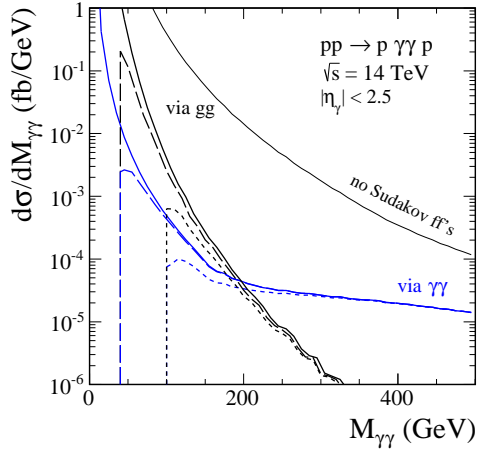}}
\end{minipage}
   \caption{
\small Distribution in invariant mass of the two-photon system for
the Durham QCD mechanism (black lines) and QED $\gamma\gamma$ fusion
mechanism (blue lines). We present results without cuts (solid line)
and with extra cuts on photon transverse
momenta $p_{\perp,\gamma} >$ 20, 50 GeV (long dashed, dashed lines,
respectively) were imposed for illustration. }
 \label{fig:backgrounds}
\end{figure*}

Up to now we have discussed cross sections and differential
distributions for technipion production in exclusive $pp$
scattering. In real experiment, an optimal decay channel must be
chosen due to presumably low production cross sections, on the one
hand, and to maximize the signal-to-background ratio, on the other
hand. In Fig.~\ref{fig:branching_fractions} we show branching
fractions for major real technipion ${\tilde \pi}^0$ decay channels.
In a very broad range of technipion and techniquark masses the two-photon
decay channel seems to be the most optimal one. In addition, this is
one of the golden channels for Higgs boson searches and the LHC
detectors are well suited for such studies.

Let us concentrate now on the exclusive diphoton background to the
exclusive technipion production. In Fig.~\ref{fig:backgrounds} we
show the corresponding distribution in invariant mass of the two
outgoing photons $M_{\gamma\gamma}$. We show distributions for the
Durham QCD mechanism and for the QED $\gamma\gamma$ fusion mechanism
calculated based upon the parton-model formula
(\ref{exclusive_gamgam}). At relatively low
masses, the Durham mechanism dominates. However, above
$M_{\gamma \gamma} >$ 200 GeV the photon-photon mechanism takes
over. The later is therefore the most important potential background
for the technipion signal if observed in the $\gamma\gamma$ decay
channel. For the pQCD background we have also shown a result without
Sudakov formfactors. As can bee seen from the figure the Sudakov formfactors
strongly damp the cross section, especially at larger photon-photon 
invariant masses.
Assuming the experimental resolution in invariant
$\gamma\gamma$ mass of about 5 GeV or so, the background turns
out to be by two orders of magnitude smaller than the corresponding
technipion signal for the whole range of vector-like TC model
parameters considered in the present paper. To summarize, the
signal-to-background ratio in exclusive technipion production
process is by far better than that in inclusive technipion
production \cite{VLTC}. The latter is clear from comparing the
corresponding inclusive $\gamma\gamma$ background estimates which
have been done earlier in the Higgs boson $\gamma\gamma$ signal
studies at the LHC \cite{ATLAS,CMS} and typical inclusive technipion
production cross sections shown e.g. in Fig.~\ref{fig:inclusive}.

In Table~\ref{tab:cross_sections} we list the total $pp\to
p(\gamma\gamma)p$ exclusive cross sections at the LHC ($\sqrt{s}=14$
TeV) for the QCD $gg\to\gamma\gamma$ and QED
$\gamma\gamma\to\gamma\gamma$ mechanisms in separate 50 GeV - windows
in diphoton $\gamma\gamma$ invariant mass $M_{\gamma\gamma}$ placed
between 50 and 400 GeV of diphoton invariant mass.
A realistic cut on both photon pseudorapidities
$|\eta_{\gamma}|<2.5$ is imposed. For comparison, we show the
numerical results with an extra cut on transverse momenta of both outgoing
photons $p_{\perp,\gamma}>$ 50 GeV and without it, as well as for
two different choices of the gluon PDFs \cite{GJR,MSTW} entering 
the calculation of UGDF in the Durham approach (c.f. Eq.~(\ref{ampl})). 
As we have already observed in Fig.~\ref{fig:backgrounds}, the QCD 
component of the exclusive $\gamma\gamma$ background dominates only 
for smaller invariant masses $M_{\gamma\gamma}\lesssim 200$ GeV, 
while for larger ones the QED mechanism becomes important.
Observation of much larger cross section in only one of the windows
than those given in Table 1 would be then a probable signal of a new
resonance (technipion). On the other hand, observation of much larger
cross section in many windows simultaneously would be a signal of new
particles appearing in loops.

\begin{table}
 \caption{ The cross sections (in fb) for photon-pair central
exclusive production at $\sqrt{s} = 14$~TeV in the photon
pseudorapidity $|\eta_{\gamma}| < 2.5$ and with cuts in $p_{\perp,
\gamma} > 50$~GeV on both outgoing photons. Different choices of
gluon PDF are used at quite small values of gluon transverse momenta
$q_{\perp, min}^{2} = 0.5$~GeV$^2$. }
 \label{tab:cross_sections}
\begin{tabular}{|c|c|c|c|c|c|c|}
\hline & \multicolumn{6}{c|}{$\sigma$ (fb) at $\sqrt{s} = 14$~TeV
and
$|\eta_{\gamma}| < 2.5$} \\
\cline{2-7} $M_{\gamma \gamma}$ & \multicolumn{2}{c|}{$\gamma \gamma
\to \gamma \gamma$} & \multicolumn{2}{c|}{$gg \to \gamma \gamma$,
{\tt GJR08VFNS NLO}} &
\multicolumn{2}{c|}{$gg \to \gamma \gamma$, {\tt MSTW08 NLO}} \\
\cline{2-7} & no cuts $p_{\perp, \gamma}$ & $p_{\perp, \gamma} >
50$~GeV & no cuts $p_{\perp, \gamma}$ & $p_{\perp, \gamma} > 50$~GeV
& no cuts $p_{\perp, \gamma}$ & $p_{\perp, \gamma} > 50$~GeV \\
\hline
\;\;50 -- 100 & $97.01 \times 10^{-3}$ & -- & 3.048 & -- & 2.752 &
--
\\
100 -- 150 & $11.62 \times 10^{-3}$&  $4.10 \times 10^{-3}$ & $62.72
\times 10^{-3}$& $22.55 \times 10^{-3}$ & $67.08 \times 10^{-3}$&
$23.20 \times 10^{-3}$
\\
150 -- 200 & \;\;$2.96 \times 10^{-3}$&   $2.01 \times 10^{-3}$ &
\;\;$5.90 \times 10^{-3}$& \;\;$4.21 \times 10^{-3}$ & \;\;$6.84
\times 10^{-3}$& \;\;$4.74 \times 10^{-3}$
\\
200 -- 250 & \;\;$1.78 \times 10^{-3}$& $1.51 \times 10^{-3}$ &
\;\;$0.95 \times 10^{-3}$& \;\;$0.79 \times 10^{-3}$ & \;\;$1.15
\times 10^{-3}$& \;\;$0.94 \times 10^{-3}$
\\
250 -- 300 & \;\;$1.44 \times 10^{-3}$& $1.34 \times 10^{-3}$ &
\;\;$0.23 \times 10^{-3}$& \;\;$0.21 \times 10^{-3}$ & \;\;$0.29
\times 10^{-3}$& \;\;$0.25 \times 10^{-3}$
\\
300 -- 350 & \;\;$1.23 \times 10^{-3}$& $1.19 \times 10^{-3}$ &
\;\;$0.06 \times 10^{-3}$& \;\;$0.05 \times 10^{-3}$ & \;\;$0.07
\times 10^{-3}$& \;\;$0.07 \times 10^{-3}$
\\
350 -- 400 & \;\;$1.06 \times 10^{-3}$& $1.05 \times 10^{-3}$ &
\;\;$0.02 \times 10^{-3}$& \;\;$0.02 \times 10^{-3}$ & \;\;$0.03
\times 10^{-3}$& \;\;$0.02 \times 10^{-3}$
\\
\hline
\end{tabular}
\end{table}

\section{Summary and conclusion}

We have made a first analysis of an interesting possibility to
search for technipions mostly decaying into two photons in exclusive
$p p \to p p \gamma \gamma$ process at the LHC. We have considered a
particularly interesting case of light technipions which do not
directly interact with gluons and quarks to the leading order, but
can interact only with SM gauge bosons. A single technipion in this
case can only be produced via a techniquark triangle loop in a
vector boson fusion channel. The latter specific properties of
physical technipions are predicted, in particular, by recently
suggested phenomenologically consistent vector-like Technicolor (TC)
model \cite{VLTC}. We have calculated the dependence of the $p p \to
p p \tilde \pi^0$ cross section on the vector-like TC model
parameters. With a natural choice of parameters obtained by a mere
QCD rescaling the corresponding cross sections of the order of one
to a few femtobarns could be expected. This means that the exclusive
${\tilde \pi}^0$ production cross section can be of the same order or
even exceeds the traditional Higgs boson CEP cross section
\cite{Durham,MPS_bbbar} making the considered proposal very
important for the forward physics program at the LHC
\cite{Albrow:2010yb,FP420}.

In the present analysis we have considered only purely exclusive
process, i.e. we have assumed that the both outgoing protons
are detected. This is not yet possible at the LHC, but could be
possible when forward proton detectors are installed by the ATLAS
and/or CMS collaborations.

We have calculated several differential distributions and discussed
their characteristic features. The particularly interesting ones are
distributions in azimuthal angle between outgoing protons. The
outgoing protons are scattered dominantly to perpendicular azimuthal
directions.

We have demonstrated that for not too large technipion masses the
photon-photon decay channel has the largest branching fraction. This
shows that the exclusive reaction $p p \to p p \gamma \gamma$ is
probably the best suited in searches for technipions at the LHC.

We have therefore studied the expected Standard Model exclusive
$\gamma\gamma$ backgrounds. We have considered two important sources
of the non-resonant background: the Durham QCD mechanism (via $gg
\to \gamma \gamma$ subprocess) and the QED mechanism (via $\gamma
\gamma \to \gamma \gamma$ subprocess). In the later case we have
included full set of box diagrams with lepton, quark and $W$ boson
loops thus focusing on the dominant Standard Model processes only.
The most interesting is the distribution in diphoton invariant mass.
At lower invariant masses, the Durham QCD mechanism dominates. At
larger invariant masses, the light-by-light rescattering occurs to
be more relevant background in searches for technipions. We conclude that the
signal-to-background ratio would be very favorable in the reaction
under consideration.

The light-by-light rescattering subprocess contribution to the
exclusive diphoton signal at the LHC in large diphoton invariant
masses is interesting in its own right as a good probe in searches
for effects beyond the Standard Model (e.g. supersymmetry, Dirac
monopoles etc). All this makes the $p p \to p p \gamma \gamma$
reaction particularly interesting for LHC phenomenology.

In the present analysis we have considered purely exclusive
processes. The related experiments would require therefore
measurements of forward protons. We hope this will be possible in a
close future \cite{FP420}. In principle, one could also allow
semi-exclusive (e.g. single diffractive) processes when excited
states of proton (proton resonances or continuum) are produced while
the pile-up problem has to be solve in high luminosity runs. The
latter will be investigated elsewhere.

\vspace{1cm} {\bf Acknowledgments}

Stimulating discussions and helpful correspondence with Johan
Bijnens, Vitaly Beylin, Vladimir Kuksa, Johan Rathsman, Francesco
Sannino, Torbj\"orn Sj\"ostrand and Grigory Vereshkov are gratefully
acknowledged. This work was supported by the Crafoord Foundation
(Grant No. 20120520) and by the Polish MNiSW grant DEC-2011/01/B/ST2/04535.
R. P. is grateful to the ``Beyond the LHC''
Program at Nordita (Stockholm) for support and hospitality during
initial stages of this work.



\begin{thebibliography}{99}

\bibitem{Albrow:2010yb}
  M.~G.~Albrow, T.~D.~Coughlin and J.~R.~Forshaw,
  Prog.\ Part.\ Nucl.\ Phys.\  {\bf 65} (2010) 149.

\bibitem{FP420}
  M.~G.~Albrow {\it et al.}  [FP420 R$\&$D Collaboration],
  JINST {\bf 4}, T10001 (2009).

\bibitem{Durham}
  V.~A.~Khoze, A.~D.~Martin and M.~G.~Ryskin,
  Phys.\ Lett.\  {\bf B401} (1997) 330;
  Eur.\ Phys.\ J.\ {\bf C14} (2000) 525;
  Eur.\ Phys.\ J.\  {\bf C19} (2001) 477 [Erratum-ibid.\  {\bf C20} (2001) 599];
  Eur.\ Phys.\ J.\  {\bf C23} (2002) 311;
\\
  A.~B.~Kaidalov, V.~A.~Khoze, A.~D.~Martin and M.~G.~Ryskin,
  Eur.\ Phys.\ J.\  {\bf C33} (2004) 261.

\bibitem{Dechambre:2011py}
  A.~Dechambre, O.~Kepka, C.~Royon and R.~Staszewski,
  Phys. Rev. {\bf D83} (2011) 054013.

\bibitem{SF}
  M.~G.~Ryskin, A.~D.~Martin, V.~A.~Khoze,
  Eur.\ Phys.\ J.\  {\bf C60} (2009) 265.

\bibitem{chic}
  R.~S.~Pasechnik, A.~Szczurek and O.~V.~Teryaev,
  Phys.\ Rev.\ D {\bf 78}, 014007 (2008); \\
  P.~Lebiedowicz, R.~Pasechnik and A.~Szczurek,
  Phys.\ Lett.\ B {\bf 701}, 434 (2011).

\bibitem{LKRS10}
  L.~A.~Harland-Lang, V.~A.~Khoze, M.~G.~Ryskin and W.~J.~Stirling,
  Eur.\ Phys.\ J.\  {\bf C69} (2010) 179.

\bibitem{LPS13}
  P.~Lebiedowicz, R.~Pasechnik and A.~Szczurek,
  Nucl.\ Phys.\ B {\bf 867}, 61 (2013).

\bibitem{Szczurek:2006bn}
  A.~Szczurek, R.~S.~Pasechnik and O.~V.~Teryaev,
  Phys.\ Rev.\ {\bf D75} (2007) 054021.

\bibitem{HarlandLang:2011qd}
  L.~A.~Harland-Lang, V.~A.~Khoze, M.~G.~Ryskin and W.~J.~Stirling,
  Eur.\ Phys.\ J.\ {\bf C71} (2011) 1714.

\bibitem{EP2011}
  R.~Enberg and R.~Pasechnik,
  Phys.\ Rev.\ {\bf D83}, 095020 (2011).

\bibitem{CMS-bump}
CMS Collaboration, CMS-PAS-HIG-13-016, 2013.


\bibitem{TC}
S.~Weinberg, Phys. Rev. {\bf D13}, 974 (1976);\\
L.~Susskind, Phys. Rev. {\bf D20}, 2619 (1979).

\bibitem{Extended-TC}
  S.~Dimopoulos and L.~Susskind,
  Nucl.\ Phys.\ {\bf B155}, 237 (1979);\\
  E.~Eichten and K.~D.~Lane,
  Phys.\ Lett.\ {\bf B90}, 125 (1980).

\bibitem{Hill:2002ap}
  C.~T.~Hill and E.~H.~Simmons,
  Phys.\ Rept.\  {\bf 381}, 235 (2003)
  [Erratum-ibid.\  {\bf 390}, 553 (2004)]
  [hep-ph/0203079].

\bibitem{Sannino}
F.~Sannino,
  Acta Phys.\ Polon.\ B {\bf 40}, 3533 (2009).

\bibitem{ATLAS}
G.~Aad {\it et al.}  [ATLAS Collaboration],
  Phys.\ Lett.\ {\bf B716}, 1 (2012);
  Science {\bf 338}, 1576 (2012).

\bibitem{CMS}
S.~Chatrchyan {\it et al.}  [CMS Collaboration],
  Phys.\ Lett.\ {\bf B716}, 30 (2012);
  Science {\bf 338}, 1569 (2012).

\bibitem{Chivukula:2011ue}
  R.~S.~Chivukula, P.~Ittisamai, E.~H.~Simmons and J.~Ren,
  Phys.\ Rev.\ D {\bf 84}, 115025 (2011)
  [Erratum-ibid.\ D {\bf 85}, 119903 (2012)].

\bibitem{Jia:2012kd}
  J.~Jia, S.~Matsuzaki and K.~Yamawaki,
  Phys.\ Rev.\ D {\bf 87}, 016006 (2013).

\bibitem{Frandsen:2012rj}
  M.~T.~Frandsen and F.~Sannino,
  arXiv:1203.3988 [hep-ph].

\bibitem{Hapola:2012wi}
  T.~Hapola, F.~Mescia, M.~Nardecchia and F.~Sannino,
  Eur.\ Phys.\ J.\ C {\bf 72}, 2063 (2012).

\bibitem{VLTC}
  R.~Pasechnik, V.~Beylin, V.~Kuksa and G.~Vereshkov,
  arXiv:1304.2081 [hep-ph].

\bibitem{VLTC-DM}
  R.~Pasechnik, V.~Beylin, V.~Kuksa and G.~Vereshkov,
  arXiv:1308.6625 [hep-ph].

\bibitem{Lee}
B.W. Lee and H.T. Nieh, Phys. Rev. {\bf 166}, 1507 (1968).

\bibitem{LSigM}
S. Gasiorowicz and D. Geffen, Rev. Mod. Phys. {\bf 41}, 531 (1969);\\
P. Ko and S. Rudaz, Phys. Rev. {\bf D 50}, 6877 (1994);\\
M. Urban, M. Buballa, and J. Wambach, Nucl. Phys. {\bf A697}, 338 (2002).

\bibitem{SU2LR}
B.D. Serot and J.D. Walecka, Acta Phys. Pol. {\bf B21}, 655 (1992).

\bibitem{ES13}
  D.~d'Enterria and G.~G.~da Silveira,
  Phys.\ Rev.\ Lett.\  {\bf 111}, 080405 (2013).

\bibitem{LS13}
P. Lebiedowicz and A. Szczurek, Phys. Rev. {\bf D87}, 074037 (2013).

\bibitem{MPS2011_gg}
R.~Maciu{\l}a, R.~Pasechnik and A.~Szczurek, Phys. Rev. {\bf D84}
114014 (2011).

\bibitem{MPS_bbbar}
  R.~Maciula, R.~Pasechnik and A.~Szczurek,
  Phys.\ Rev.\ {\bf D82}, 114011 (2010);
\\
  Phys. Rev. {\bf D83}, 114034 (2011).

\bibitem{LS2010}
P. Lebiedowicz and A. Szczurek, Phys. Rev. {\bf D81}, 036003 (2010).

\bibitem{FC}
  T.~Hahn,
  Comput.\ Phys.\ Commun.\  {\bf 140}, 418 (2001);
\\
  T.~Hahn and M.~Perez-Victoria,
  Comput.\ Phys.\ Commun.\  {\bf 118}, 153 (1999);
\\
  T.~Hahn,
  Comput.\ Phys.\ Commun.\  {\bf 178}, 217 (2008).

\bibitem{JT94}
  G.~Jikia and A.~Tkabladze,
  Phys.\ Lett.\ B {\bf 323}, 453 (1994);\\
  G.~J.~Gounaris, P.~I.~Porfyriadis and F.~M.~Renard,
  Eur.\ Phys.\ J.\ {\bf C9}, 673 (1999);
  Phys.\ Lett.\ B {\bf 452}, 76 (1999)
  [Erratum-ibid.\ B {\bf 513}, 431 (2001)].

\bibitem{Dixon}
  Z.~Bern, A.~De Freitas, L.~J.~Dixon, A.~Ghinculov and H.~L.~Wong,
  JHEP {\bf 0111}, 031 (2001).

\bibitem{DZ89}
M. Drees and D. Zeppenfeld, Phys. Rev. {\bf D39}, 2536 (1989).

\bibitem{OWZ94}
J. Ohnemus, T. F. Walsh and P. M. Zerwas, Phys. Lett. {\bf B328},
369 (1994).

\bibitem{GS98}
I. F. Ginzburg and A. Schiller, Phys. Rev. {\bf D57}, 6599 (1998).

\bibitem{D99}
H. Davoudiasi, Phys. Rev. {\bf D60}, 084022 (1999).

\bibitem{Ch00}
K.-M. Cheung, Phys. Rev. {\bf D61}, 015005 (2000).

\bibitem{Lai:1999wy}
  H.~L.~Lai {\it et al.}  [CTEQ Collaboration],
  Eur.\ Phys.\ J.\ C {\bf 12}, 375 (2000).

\bibitem{GJR}
M. Gl\"uck, P. Jimenez-Delgado, E. Reya,  
Phys. Lett. {\bf B664} 133 (2008).

\bibitem{MSTW}
A.D. Martin, W.J. Stirling, R.S. Thorne, G. Watt, 
Eur. Phys. J. {\bf C63} 182 (2009).

\end{thebibliography}
\end{document}